\documentclass[twocolumn,prl,prx,notitlepage,superscriptaddress, reprint]{revtex4}

\usepackage{graphics,epsfig,amsfonts,amssymb,amsmath,color}
\usepackage[normalem]{ulem}
\usepackage{bm,bbm}
\usepackage[mathcal]{euscript}
\usepackage[monochrome]{xcolor}
\usepackage{standalone}
\usepackage{amsmath}
\usepackage{amsthm}

\makeatletter
\def\maketitle{
	\@author@finish
	\title@column\titleblock@produce
	\suppressfloats[t]}
\makeatother

\usepackage{float}
\newcommand{\bra}[1]{\ensuremath{\langle{#1}|}}
\newcommand{\ket}[1]{\ensuremath{|{#1}\rangle}}

\newcommand{\red}[1]{\textcolor{red}{#1}}

\begin{document}

\title{Two-dimensional non-Hermitian skin effect in an ultracold Fermi gas}

\author{Entong Zhao}
\thanks{These authors contributed equally to this work.}
\affiliation{Department of Physics, The Hong Kong University of Science and Technology,\\ Clear Water Bay, Kowloon, Hong Kong, China}

\author{Zhiyuan Wang}
\thanks{These authors contributed equally to this work.}
\affiliation{International Center for Quantum Materials, School of Physics, Peking University, Beijing 100871, China}

\author{Chengdong He}
\affiliation{Department of Physics, The Hong Kong University of Science and Technology,\\ Clear Water Bay, Kowloon, Hong Kong, China}

\author{Ting Fung Jeffrey Poon}
\affiliation{International Center for Quantum Materials, School of Physics, Peking University, Beijing 100871, China}

\author{Ka Kwan Pak}
\affiliation{Department of Physics, The Hong Kong University of Science and Technology,\\ Clear Water Bay, Kowloon, Hong Kong, China}

\author{Yu-Jun Liu}
\affiliation{Department of Physics, The Hong Kong University of Science and Technology,\\ Clear Water Bay, Kowloon, Hong Kong, China}

\author{Peng Ren}
\affiliation{Department of Physics, The Hong Kong University of Science and Technology,\\ Clear Water Bay, Kowloon, Hong Kong, China}

\author{Xiong-Jun Liu}
\email{xiongjunliu@pku.edu.cn}
\affiliation{International Center for Quantum Materials, School of Physics, Peking University, Beijing 100871, China}
\affiliation{Hefei National Laboratory, Hefei 230088, China}
\affiliation{International Quantum Academy, Shenzhen 518048, China}

\author{Gyu-Boong Jo}
\email{gbjo@ust.hk}
\affiliation{Department of Physics, The Hong Kong University of Science and Technology,\\ Clear Water Bay, Kowloon, Hong Kong, China}
\affiliation{IAS Center for Quantum Technologies, The Hong Kong University of Science and Technology, Hong Kong, China}

\begin{abstract}	
 {The concept of non-Hermiticity has expanded the understanding of band topology leading to the emergence of counter-intuitive phenomena. One example is the non-Hermitian skin effect (NHSE), which involves the concentration of eigenstates at the boundary. However, despite the potential insights that can be gained from high-dimensional non-Hermitian quantum systems in areas like curved space, high-order topological phases, and black holes, the realization of this effect in high dimensions remains unexplored. \red{Here,  we create a two-dimensional (2D) non-Hermitian topological band for ultracold fermions in spin-orbit-coupled optical lattices with tunable dissipation, which exhibits the NHSE. We first experimentally demonstrate pronounced nonzero spectral winding numbers in the complex energy plane with non-zero dissipation, which establishes the existence of 2D skin effect. Further, we observe the real-space dynamical signature of NHSE in real space by monitoring the center of mass motion of atoms.}
 Finally, we also demonstrate that a pair of exceptional points (EPs) are created in the momentum space, connected by an open-ended bulk Fermi arc, in contrast to closed loops found in Hermitian systems. The associated EPs emerge and shift with increasing dissipation, leading to the formation of the Fermi arc. Our work sets the stage for further investigation into simulating non-Hermitian physics in high dimensions and paves the way for understanding the interplay of quantum statistics with NHSE.}
\end{abstract}

\maketitle

{Hermiticity of a Hamiltonian that guarantees the conserved energy with real eigenvalues breaks down when a quantum system exchanges particles and energy with the environment~\cite{ashida2020non}. This open quantum system may be effectively described by a non-Hermitian Hamiltonian which is indeed a ubiquitous description of classical systems with gain and loss~\cite{el2018non, helbig2020generalized, ezawa2019non, Zhou.2018, ghatak2020observation, liu2021non}, interacting electronic systems~\cite{Kozii.2017,Shen.2018} and optical quantum gas~\cite{ozturk2021observation}. Recently, the non-Hermitian concept has been further generalized to the periodic lattice system in which non-Hermiticity interplays with band topology and profoundly changes the band structure leading to novel features such as a bulk open-ended Fermi arc connecting two exceptional points (EPs) showing a half-integer topological charge~\cite{Zhou.2018,Bergholtz.2021}.

One of the intriguing phenomena in non-Hermitian bands is the non-Hermitian skin effect (NHSE)~\cite{yao2018edge, Yao.2018m9, Kunst.2018, Yokomizo.2019,guo2021theoretical,zhou2021engineering, Li.2022hyn}, involving the accumulation of eigenstates at the boundary of an open system. It has been noted that the conventional Bloch band theory breaks down with non-Hermiticity~\cite{yao2018edge, Yao.2018m9, Kunst.2018, Yokomizo.2019,guo2021theoretical,zhou2021engineering, Li.2022hyn}, and the exceptional degeneracy has been pointed out as a precursor of the NHSE~\cite{Kawabata.2019,Zhang.2022}. Numerous studies revisit the bulk-boundary correspondence by considering the interplay between band topology, symmetry, and NHSE, but most previous works have focused on the 1D configuration~{\cite{liang2022dynamic, weidemann2020topological, xiao2020non}}. Although the theoretical framework for understanding NHSE is mostly well-established in 1D\red{~\cite{Yokomizo.2019,sato2020origin,GBZ2020}}, the NHSE in higher dimensions or with extra physical degrees of freedom - e.g. symmetry, long-range coupling, and band topology - are still limited both in theory and experiment~\cite{hofmann2020reciprocal, zhang2021observation, zou2021observation, shang2022experimental, zhang2021observationOf,Li.2022hyn,zhou2023observation,wang2023experimental,wan2023observation}.
Moreover, the higher-dimensional NHSE interplays with or could be mapped to various fundamental Hermitian scenarios in, e.g. a curved space~\cite{duality2022, Hermitization2022, zhou2022curve}, black holes~\cite{spinblackhole2012, Weylblackhole2022}, quantum information~\cite{entanglement2021,Anomaly2021,entangletrans2023}, and higher-order topological phases~\cite{Yokomizo2020second,highorder2020}, which all necessitate the realization in many-body systems beyond one-dimension (1D).} Experimentally, however, the simultaneous realization of exceptional degeneracy and the NHSE still remains to be a formidable challenge, especially unexplored in quantum systems, nor in higher dimensions. In particular, no fermionic system has been experimentally realized with non-Hermitian topological band.


\begin{figure*}[htbp]
	\includegraphics[width=0.8\linewidth]{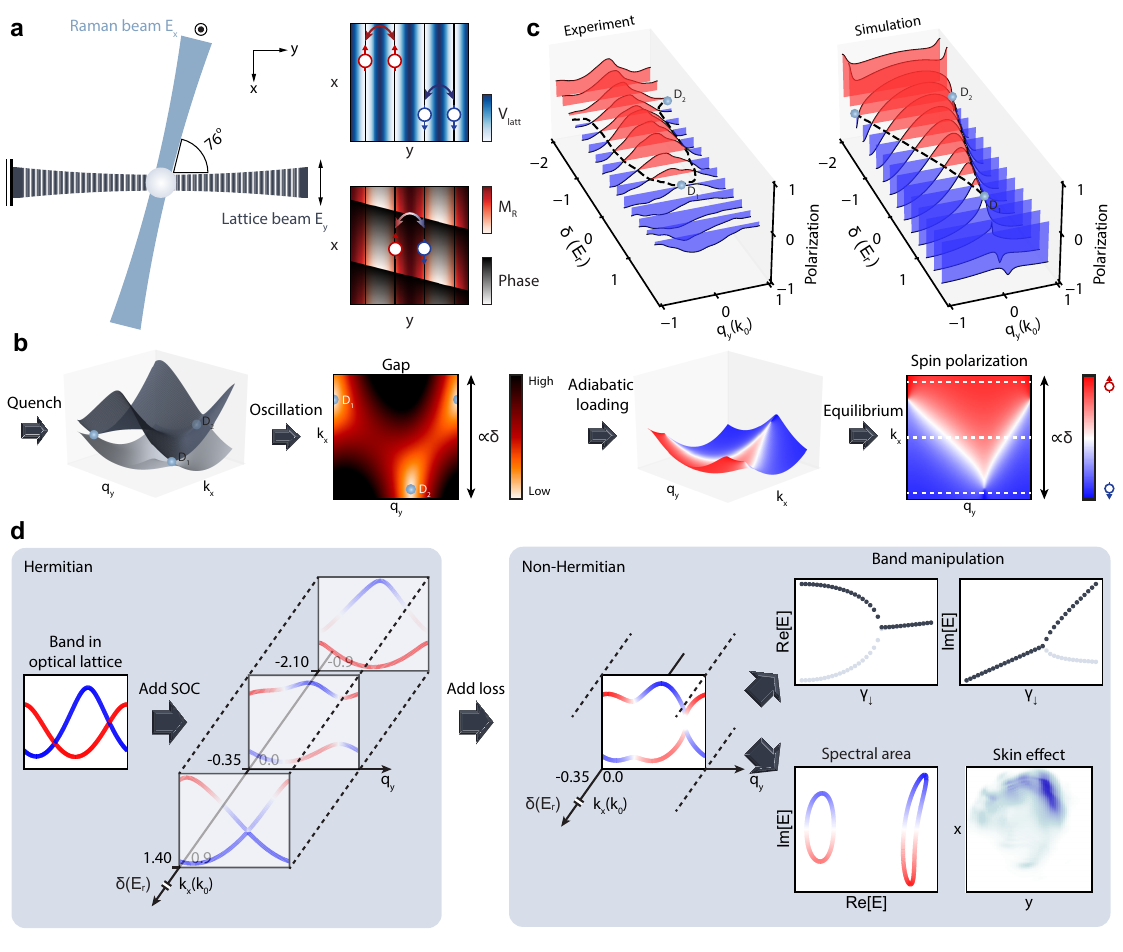}
	\caption{\textbf{\red{Hermitian and} non-Hermitian system in an optical lattice with spin-orbit coupling} \textbf{a}, Our optical Raman lattice system, which consists of a 1D optical lattice along the y direction and another Raman beam along the direction with angle $76^o$ tilted from the lattice incident beam, generating a periodic Raman potential. A loss beam with $\sigma^-$ polarization is further applied along -z direction to realize the non-Hermiticity. \textbf{b}, Band structure after the quantum quench where the band gap can be extracted by the time evolution of spin polarization and the equilibrium spin polarization which can be obtained by adiabatically loading the atoms into the lowest energy band \red{in the Hermitian optical Raman lattice}. \textbf{c}, Experimental measurement and numerical simulation of spin polarization of equilibrium state as a function of two photon detuning $\delta$. \textbf{d}, Band structure of the lowest two bands at different two-photon detuning values \red{in the Hermitian optical Raman lattice}. With increasing dissipation, atoms accumulate at the boundary of the system showing the non-trivial winding of eigenenergy in the complex energy plane. Furthermore, the energy band closes the gap at a specific quasi-momentum.}\label{fig:1}
\end{figure*}

 Here, we realize 2D NHSE for a non-Hermitian topological band in ultracold Fermi gas by synthesizing dissipative spin-orbit coupling (SOC) in a Raman-dressed lattice~\cite{song2018observation, liu2013manipulating}.  This platform allows observing both EPs and NHSE with dissipation implemented by the spin-selective atom loss~\cite{ren2022chiral}. With increasing dissipation, the Dirac point, where two bands linearly intersect, extends forming a bulk Fermi arc that connects two exceptional points. With the momentum-dependent Rabi spectroscopy, we probe the change of band gap versus dissipation strength and also the parity-time ($\mathcal{PT}$)  symmetry-breaking transition across the exceptional points. \red{Evidence from both the direct measurement of spectral topology in the complex energy plane~\cite{Zhang.2020k3c,Zhang.2022} and the observation of the dynamical signature of NHSE in real space supports the existence of a 2D skin effect being consistent with theoretical calculations.} Our work makes also a step in experimentally studying how the band topology interplays with symmetry and non-Hermiticity in high dimensions.  
\vspace{5pt}
\paragraph{\bf Tunable non-Hermitian topological band} We generate a non-Hermitian topological band by combining a simple optical lattice along the y direction with a periodic Raman potential over the x-y plane as described in Fig.~\ref{fig:1}a~\cite{Liu.2014}. Non-Hermiticity arises from the state-dependent atom loss, which enables controlled dissipation with the near-resonant optical transition. In the Hermitian regime without dissipation, a topological band in the y direction is spanned out over the x direction due to the momentum shift \red{and spin-flip transition driven} by the Raman potential~\cite{song2019observation}. As a result, two gapless Dirac points (indicated by $D_1,D_2$ in Fig.~\ref{fig:1}b) emerge in two ends of band inversion line where \red{spin-up and spin-down subbands cross and thus} are resonantly coupled through SOC (see Fig.~\ref{fig:1}b \red{and Fig.~\ref{fig:1}c}). The spin texture and relevant topological invariant can be directly measured by loading atoms adiabatically into the lowest energy band exhibiting non-trivial spin textures~{\cite{song2018observation}} (See Supplementary Information and Fig.~\ref{fig:1}c).

In the non-Hermitian regime with dissipation on, however, the complex energy spectrum necessitates the spectroscopic measurement of the complex energy gap to identify real and imaginary parts of energy gap, corresponding to band gap and damping rate, respectively. The energy spectrum within the momentum space also allows us to obtain the {\it spectral topology} of the system, leading to the observation of the NHSE~\cite{ yao2018edge, Yao.2018m9, Kunst.2018, Yokomizo.2019,guo2021theoretical}. In the current work, we probe the quench evolution of spin polarization within the energy band and elucidate non-Hermitian phenomena including the emergence of exceptional points and NHSE (See Fig.~\ref{fig:1}d).

\begin{figure*}[htbp]
	\includegraphics[width=0.65\linewidth]{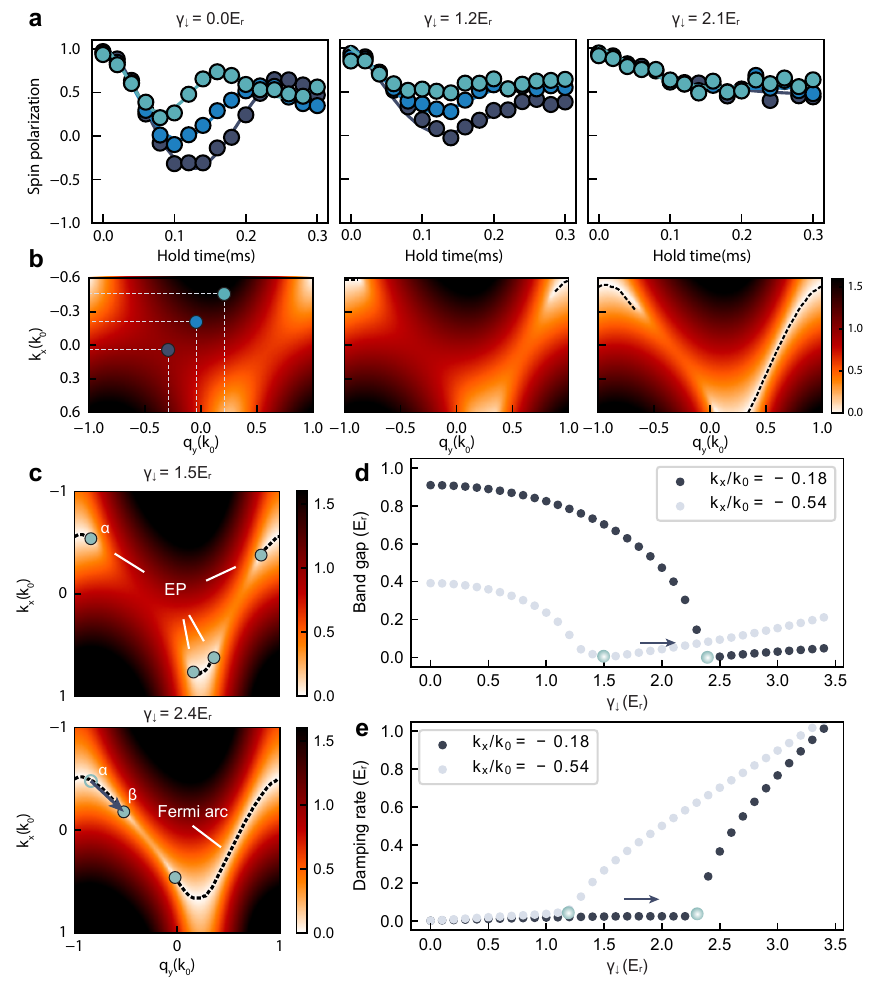}
	\caption{\textbf{Momentum-dependent Rabi oscillation to resolve the band gap closing.} \textbf{a}, Rabi oscillation at different quasi-momenta $\textbf{q}=(k_x,q_y)$ (points in \textbf{b}). Lines show the fitted curve with a damped sinusoidal function. \textbf{b}, Band gap between the lowest two bands simulated from plane wave expansion at different loss terms from $\gamma_{\downarrow}=0E_r$ to $2.1E_r$, while the lattice potential and Raman coupling strength are fixed to $V_{\uparrow(\downarrow)}=3.2(2.2)E_r$ and $M_0=2.0E_r$. The two-photon detuning $\delta$ (or equivalently $m_z$) is set to $\delta=-0.35E_r$ to compensate the on-site energy difference induced by spin-dependent lattice potential. The dashed line marks the positions of quasi-momentum $\textbf{q}$ where the energy gap is closed.   \textbf{c}, With dissipation, each Dirac point in the Hermitian regime separates into a pair of EPs, which are connected by a (non-Hermitian) Fermi arc and continuously move in the momentum space as dissipation increases. \textbf{d,e} The change of the real energy band gap and damping rate at different quasi-momenta shown by the point $\alpha$ and $\beta$ in \textbf{c} with increasing dissipation. The slight increase in the real band gap when the loss rate is larger than the exceptional point is a result of the higher band effect.}\label{fig:2}
\end{figure*}

The total Hamiltonian realized experimentally reads 
\begin{equation}
\begin{aligned}
H_0&=\sum_{\sigma=\uparrow,\downarrow}\left[\frac{p^2_x+p^2_y}{2m}+V_\sigma(y)|\sigma\rangle\langle\sigma|\right]+V_R(x,y)\\
&+m_z(\ket{\uparrow}\bra{\uparrow}-\ket{\downarrow}\bra{\downarrow})+H_{loss}
\end{aligned}\label{expmodel}
\end{equation}
\noindent where $\frac{p_x^2}{2m}$~($\frac{p_y^2}{2m}$) is the kinetic energy term in the x~(y) direction, $V_R=M_R(x,y)\ket{\downarrow}\bra{\uparrow}+\text{h.c.}$ is the Raman coupling term and $H_{loss}=-\sum_{\sigma=\uparrow,\downarrow}\frac{i}{2}\gamma_{\sigma}\ket{\sigma}\bra{\sigma}$ represents spin-sensitive atom loss. The Zeeman term $m_z=(\delta-\delta_0)/2$ can be controlled by two photon detuning $\delta$, where $\delta_0$ denotes the on-site energy difference between $\bra{\uparrow}$ and $\ket{\downarrow}$ states. When there is no spin-selective dissipation ($\gamma_{\downarrow (\uparrow)}=0$), the system is Hermitian and the Raman coupling opens the band gap around the band-inversion line (Fig.~\ref{fig:1}d), \red{which is the particular quasi-momentum subspace with resonant spin flipping ~\cite{liang2021realization,song2019observation}}.  \red{The Raman potential can be describe by $M_R(x,y)=M_0\cos(k_0y)e^{ik_2x-ik_1y}$ with $k_1=k_0\sin\theta$ and $k_2=k_0\cos\theta$, and $e^{ik_2x}$ term} in Raman potential coupling the two spin states gives the x-directional kinetic energy $\frac{\hbar^2}{2M}(k_x\pm\frac{k_2}{2})^2$, rendering the linear spin-orbit term $\frac{\hbar^2}{2M}(k_xk_2)\sigma_z$, which leads to a 2D energy band (see Fig.~\ref{fig:1}b). In fact, the linear spin-orbit term effectively shifts the Zeeman energy $m_z$, which makes the energy band for the $k_x=0$ layer obtained by tuning $m_z$ is identical to that for fixed $m_z=m_0$ by scanning $k_x$ following the relation $H_{q_y,k_x=0,m_z=m_0}=H_{q_y,k_x=\frac{2Mm_0}{\hbar^2k_2},m_z=0}$.

\begin{figure*}[htbp]
	\includegraphics[width=0.8\linewidth]{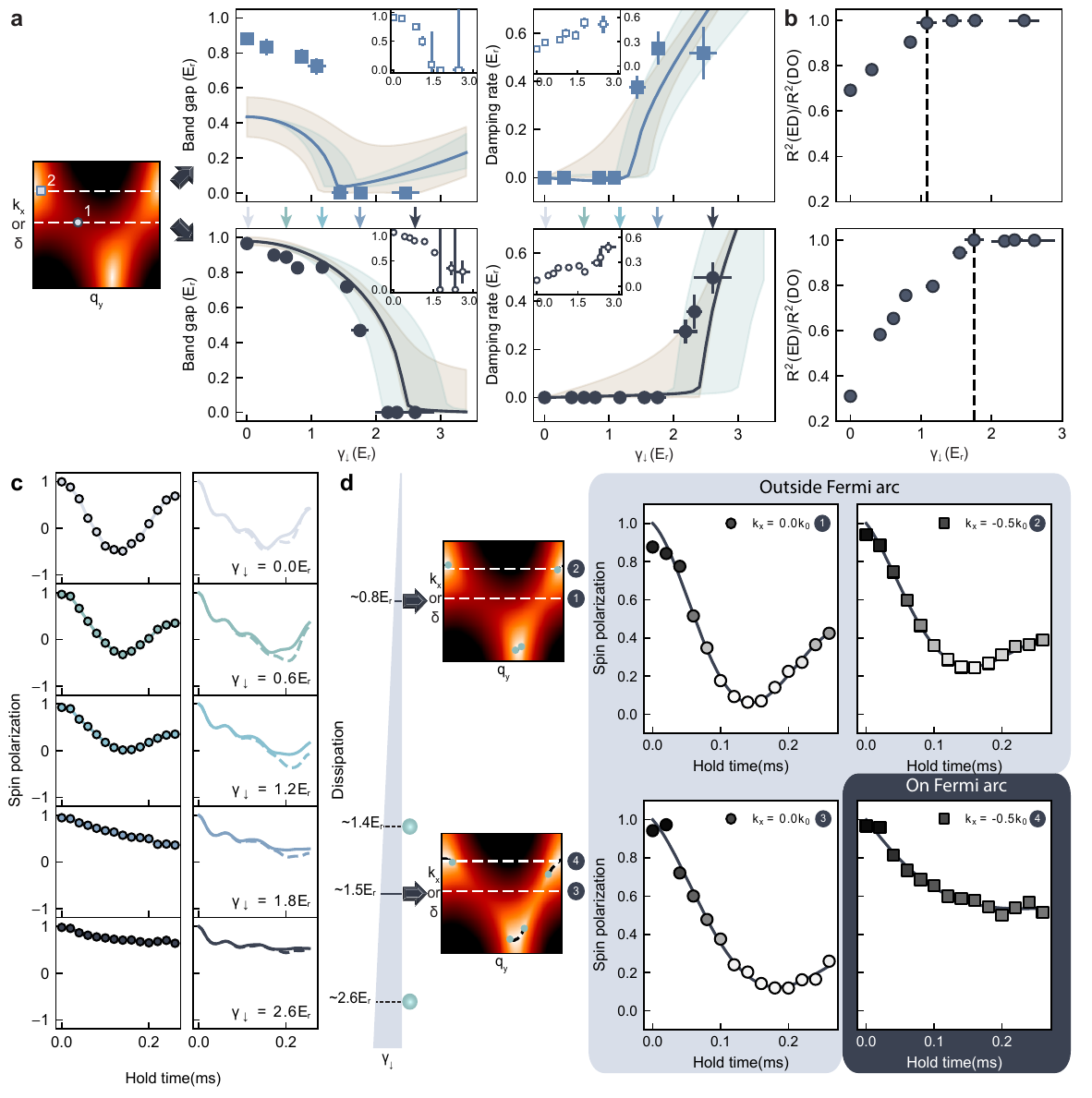}
		\caption{{\bf Observation of non-Hermitian EPs and Fermi arc.} \textbf{a} The band gap $\Re{(\Delta E)}$ (left) and damping rate $\Im{(\Delta E)}$ (right) extracted from the Rabi oscillation at two different quasi-momentum marked in the left energy band gap. Before and after the EPs, the fitting functions are different. The solid line is obtained from the simulation of plane wave expansion. The shaded region indicates the uncertainty of the band gap and damping rate associated with the uncertainty of atom loss (green color) and the uncertainty of quasi-momentum (brown color). The inset shows the fitting result with same damped oscillation fitting function before and after the exceptional point. \red{The vertical and horizontal error bars represent the fitting errors and the experimental uncertainty related to the calibration, respectively.} \textbf{b} The normalized $R^2$ value, obtained by fitting the experimental data with exponential decay (ED) curves, divided by the $R^2$ value obtained from fitting with a damped oscillation (DO) curve, is presented. The corresponding quasi-momentum values, $(k_x,q_y)=(-0.5,-0.9)k_0$ (top) and $(k_x,q_y)=(0.0,-0.3)k_0$ (bottom), are marked in the energy band gap shown in \textbf{a}, respectively. This ratio reveals that above the exceptional point (EP), the fitting using an exponential decay function performs equally well as the fitting with a damped oscillation, indicating that the spin oscillation follows a monotonic exponential behavior, while below the EP, the exponential decay function is disfavored. \red{The horizontal error bars represent the experimental uncertainty related to the calibration.} \textbf{c} The spin oscillation for same quasi-momentum and different dissipation strengths and marked by the arrows in \textbf{a} bottom. When $\gamma_\downarrow$ is smaller than the EP, the spin polarization oscillates with the frequency dependent, while above EP, the PT-symmetry broken phase shows a monotonic spin polarization. The right column shows the corresponding simulation from plane wave expansion with finite temperature considering momentum uncertainty (solid line) and without considering the momentum uncertainty (dashed line). \textbf{d} The spin oscillation for two different quasi-momentum and similar dissipation strength $\gamma_\downarrow\sim0.8E_r$ (top) and $1.5E_r$ (bottom). When $\gamma_{\downarrow}=1.5E_r$, the left EP moves across $k_x=-0.5k_0$, which means the quasi momentum $(k_x,q_y)=(-0.5,-0.9)k_0$ is on Fermi arc and shows a monotonic spin polarization. }\label{fig:3}
\end{figure*}

\paragraph{\bf Probing quench evolution in the non-Hermitian regime} To monitor quench evolution of non-Hermitian topological band, we begin with preparing a spin-polarized degenerate $^{173}$Yb Fermi gas in the spin-up states $\ket{\uparrow}$ followed by loading into a Hermitian optical lattice potential within 3~ms (See Fig.~\ref{fig:1}b and d). At this stage, two-photon detuning $\delta_i\sim-27E_r$ is set to be far enough to minimize SOC where $E_r=h\times 3.735$~kHz is the recoil energy. After the initial-state preparation with additional 1~ms hold, two photon-detuning is suddenly adjusted from $\delta_i$ to $\delta_f$ with the loss pulse being switched on at time $t=0$ (Fig.~\ref{fig:1}d). The band structure of the system starts to evolve from a hermitian lattice band to a band with dissipative SOC, and the atoms initially in $\ket{\uparrow}$ state start the Raman-Rabi oscillation between $\ket{\uparrow}$ and $\ket{\downarrow}$. Holding a variable time $t_f$, we record spin-sensitive momentum distribution after 10ms (or 15ms) time-of-flight expansion and extract the oscillation of spin polarization in the first Brillouin zone. In Fig.\ref{fig:2}a, we show a typical spin oscillation in the momentum space after the quench with increasing dissipation $\gamma_{\downarrow}$. The spin polarization at different quasi-momentum oscillates with different frequencies revealing a non-uniform band gap in consistent with the calculated band gap (see Fig.~\ref{fig:2}b). 


\begin{figure*}
	\includegraphics[width=1.0\linewidth]{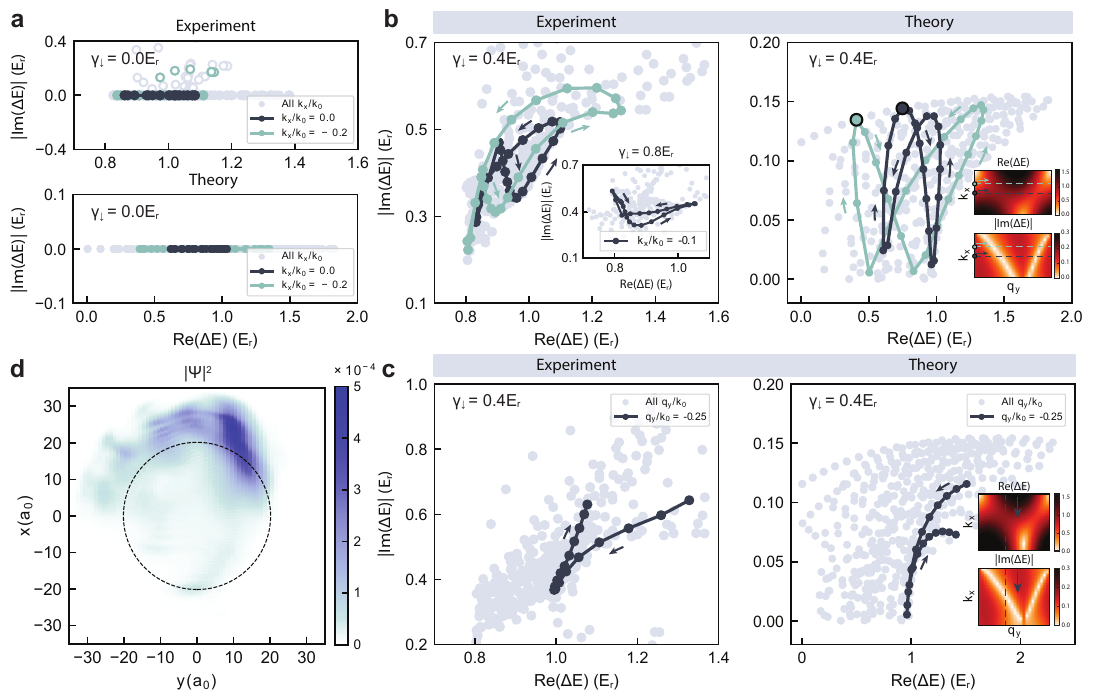}
	\caption{\textbf{Signature of NHSE \red{in momentum space}.} \textbf{a}, Experimental and simulated band difference along $q_y$ direction without dissipation ($\gamma_{\downarrow}=0.0E_r$) on the complex plane. The real part is extracted from spin polarization curve, while the imaginary part is extracted from spin down curve. \textbf{b} Experimental and simulated band difference along $q_y$ direction with dissipation strength $\gamma_{\downarrow}=0.4E_r$ on the complex plane. The dots correspond to the difference between the lowest two eigenvalues at all quasi-momentum within first Brillouin zone and $-0.6<k_x/k_0<0.6$. The solid lines correspond to the spectral flow of the band gap at different $k_x$ layer when $q_y$ runs from $-k_0$ to $k_0$. The inset in the left figure shows band difference on the complex plane when spin-selective dissipation $\gamma_{\downarrow}=0.8E_r$. The inset in the right figure shows the real and imaginary band gap at $\gamma_{\downarrow}=0.4E_r$ simulated from the plane wave expansion. \textbf{c} Experimental and simulated band difference along $k_x$ direction on the complex plane with the dissipation strength  $\gamma_\downarrow=0.4E_r$. The dots correspond to the difference between the lowest two eigenvalues at all quasi-momentum within first Brillouin zone and $-0.6<k_x/k_0<1.0$. The solid lines correspond to the spectral flow of the band gap at selected $q_y$ layer labeled by arrow in inset.  \textbf{d}, Simulation of density profile \red{in the harmonic trap} with $\gamma_{\downarrow}=0.4$$E_r$ \red{and trap frequency $\omega_r=2\pi\times112$Hz}, where $a_0$ is the lattice constant and the finite-difference step-length in $x$ direction is $x_0=a_0/7.0$. The density profile is summed over eigenstates with eigenenergy satisfying $\Re(E)<1.5E_r$. \red{The dashed line denotes the effective trapping size obtained from the Fermi energy.}}\label{fig:4}
\end{figure*}

\paragraph{\bf Emergence of non-Hermitian Fermi arc with exceptional points}
One key feature arising from the appearance of dissipation in our system is the emergence of exceptional degeneracy forming a non-Hermitian Fermi arc (connecting two exceptional points), which affects the physical properties of energy band and reveals non-trivial band topology~\red{\cite{zhou2021engineering, zhen2015spawning, Kozii.2017}}. In the Hermitian band, the band gap near the band inversion line is opened by the SOC (see Fig.~\ref{fig:1}d) while the band gap becomes smaller with increasing dissipation due to the competition between SOC and dissipation as described in Fig.~\ref{fig:2}b~\cite{ren2022chiral}. With increasing dissipation, two exceptional points, at which two eigenstates coalesce, emerge and continuously move in the momentum space, being connected by a (non-Hermitian) Fermi arc denoted by the dashed line (see Fig.~\ref{fig:2}c). We note that the non-Hermitian Fermi arc is a bulk phenomenon in contrast to the Hermitian counterpart, the surface Fermi arc in 3D Weyl semimetal. To reveal a Fermi arc with exceptional points, we examine the spin evolution after quenching at two quasimomenta $\alpha$=$(k_x, q_y)\simeq(-0.54, -0.84)k_0$ and $\beta$=$(k_x, q_y)\simeq(-0.18, -0.54)k_0$ on the band inversion line. As the dissipation strength increases, the EP moves to the point $\alpha$ closing the band gap while the point $\beta$ still opens the gap. At larger dissipation strength, the EP reaches the point $\beta$ while the quansimomentum between $\alpha$ and $\beta$ closes the gap with a non-zero damping rate (Fig.~\ref{fig:2}d). Notably, the real band gap does not remain zero after EP, as the higher bands cause a slight shift in the Fermi arc with increasing dissipation. We also notice there is no such effect in the simulation using the tight-binding model.

\paragraph{\bf Observation of the EPs and Fermi arc} Fig.~\ref{fig:3} shows the measurement of the band gap $\Re{(\Delta E)}$ and damping rate $\Im{(\Delta E)}$ at different momentum states 1 and 2 for variable dissipation $\gamma_{\downarrow}$. Our measurement manifests the EP moves in the momentum space passing the point 2 and 1, consecutively, showing the growth of the Fermi arc with increasing dissipation. The energy band gap before EP and damping rate after EP can then be extracted by fitting the oscillation curve with a sinusoidal and a monotonic exponential function, respectively (see Methods). By gradually increasing the dissipation, we observe that above the EP, the spin oscillation exhibits a monotonic exponential behavior, while below the EP, a monotonic exponential function clearly fails to capture the observed dynamics. This observation is quantitatively assessed by extracting the $R^2$ values of the exponential decay function and the damped oscillation function at different dissipation strengths. The resulting ratios are calculated and depicted in Fig.~\ref{fig:3}b, providing an alternative approach to probe the appearance of the EP. In the inset, the result obtained by fitting only with a damped oscillation function is also presented. It should be noted that the band gap cannot completely reach zero due to the limitations imposed by the finite momentum resolution and fitting constraints.

When the loss rate is smaller than the critical dissipation, the initial state at point 1 oscillates between two eigenstates with the minimal damping rate respecting the $\mathcal{PT}$ symmetry, as consistent with calculations (Fig.~\ref{fig:3}c). Beyond the critical value, however, the strong dissipation completely closes the band gap making time evolution of spin polarization exponentially decay, manifesting $\mathcal{PT}$ symmetry breaking. In Fig.~3d, we show the comparison of the spin polarization at two different quasi momentum $(k_x,q_y)=(-0.5,-0.9)k_0$ and $(k_x,q_y)=(0.0,-0.3)k_0$ with similar dissipation strength $\gamma_{\downarrow}\sim0.8E_r$ and $1.5E_r$.  At a smaller dissipation strength, the spin polarization at both quasi-momentums shows oscillation behaviors, which means they are outside the Fermi arc. With increasing dissipation, the spin polarization for higher $|k_x|$ convert into a monotonic curve earlier than lower $|k_x|$, which is another proof of the growth of Fermi arc. 

With the above results we have successfully demonstrated the realization of a 2D non-Hermitain topological band described by Hamiltonian in Eq.~\eqref{expmodel}. The Hermitian nature of this Hamiltonian is confirmed through measurements of spin polarization in the equilibrium state or quench dynamics (see Supplementary Information), and our investigation extends the understanding to the non-Hermitian regime with the observation of EPs and Fermi arc phenomena. In addition to these remarkable findings, our platform also enables the observation of the NHSE arising from the dissipation.

\begin{figure*}
	\includegraphics[width=0.85\linewidth]{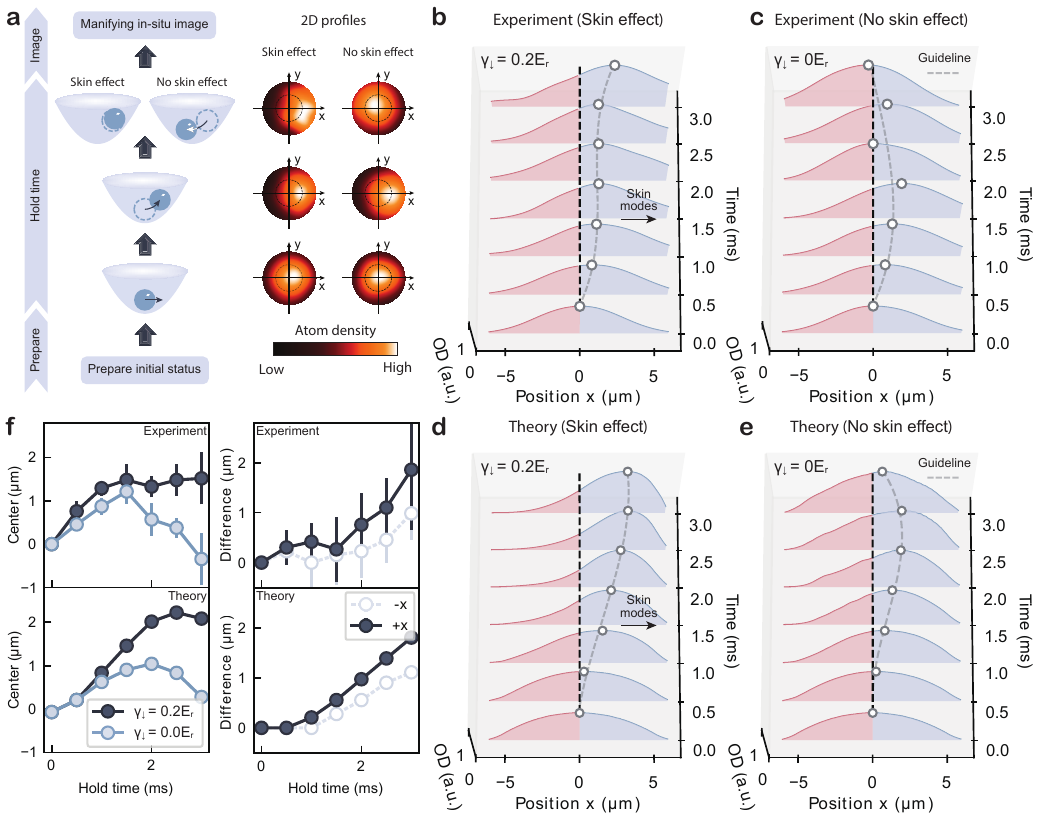}
	\caption{\red{\textbf{Signature of NHSE in real space.} \textbf{a}, Schematic illustrating the detection of the signature of the NHSE in real space. Initially, atoms are prepared with an initial status and given an initial velocity in the +x boundary. Subsequently, the system is either quenched to a non-Hermitian Hamiltonian or conserved in a Hermitian Hamiltonian. After a certain hold time, the presence of the skin effect causes the atoms to accumulate in the +x direction. In contrast, in the absence of the skin effect, the atoms undergo dipole oscillation. \textbf{b,c} One example of experimental one-dimensional normalized spin up atom profiles along the $x$ direction with ($\gamma_{\downarrow}=0.2E_r$) or without skin effect ($\gamma_{\downarrow}=0.0E_r$). The initial motion is along $+x$ direction. The normalized atom profiles are averaged for the whole atom cloud along $y$ direction. The positions with the highest optical density (OD) are denoted by empty circles and the gray dashed line is a guideline based on polynomial function. The profiles corresponding to positive and negative positions are indicated by blue and red colors, respectively. \textbf{d,e} Simulated one-dimensional normalized spin up atom profiles along the $x$ direction with or without skin effect, specifically $\gamma_{\downarrow}=0.2E_r$, or $\gamma_{\downarrow}=0.0E_r$. The initial motion is along $+x$ direction. The normalized atom profiles are averaged for the whole atom cloud along $y$ direction. The positions with the highest optical density (OD) are denoted by empty circles and the gray dashed line is a guideline based on polynomial function. The profiles corresponding to positive and negative positions are indicated by blue and red colors, respectively. \textbf{f}, Extracted center positions (left) of the experimental and simulated spin-up atom profiles along the $x$ direction when the atom cloud has an initial velocity in the +x direction, comparing the cases with ($\gamma_{\downarrow}\sim0.2E_r$) and without the NHSE ($\gamma_{\downarrow}=0.0E_r$). The difference in center positions between the two cases with skin effect ($\gamma_{\downarrow}\sim0.2E_r$) and without skin effect when the atom cloud has an initial velocity to +x or -x direction, is shown on the right. The error bars indicate the standard deviation obtained from multiple measurements.}}\label{fig:5}
\end{figure*}

\paragraph{\bf Non-Hermitian skin effect} 
The existence of exceptional degeneracy in the spectra signals the NHSE~\red{\cite{Zhang.2022}}. Also, if the spin-dependent dissipation is projected onto the spin-orbit-coupled bands of the Hermitian part of the Hamiltonian, asymmetric hoppings will arise and lead to the NHSE~\red{\cite{yi2020non}}. To concretely characterize the NHSE in our system, we evaluate the topological feature of the complex spectrum. 
It is shown that the NHSE arises in 1D under open-boundary condition (OBC), if and only if the periodic-boundary condition (PBC) spectrum of the Hamiltonian is point-gapped topological~\red{\cite{sato2020origin, Zhang.2020k3c}}, namely having nonzero spectral winding numbers or being non-backstepping curve(s) (see Supplementary Information for details). {It can be understood that the non-backstepping spectrum under PBC indicates the lack of two degenerate Bloch waves to be superimposed to satisfy OBC, and this breakdown of the Bloch theorem leads to NHSE~\cite{Yokomizo.2019}.} In a 2D system, a nonzero periodic-boundary spectral area indicates that the NHSE persists unless the open boundary geometry coincides with certain spatial symmetries of the bulk~\red{\cite{Zhang.2022,wang2022amoeba}}. Specifically, nontrivial spectral topology of quasi-1D subsystems perpendicular to the edges can reveal the spectral flow and serve as an indication of the 2D skin effect under rectangular open boundary geometry~\red{\cite{Zhang.2022,wang2022amoeba,zhang2023edge}}.

{In our experiment, we investigate the existence of the NHSE with spectral topology of complex band difference extracted from the quench evolution of spin polarization and spin down information. If there is no dissipation, the experimentally measured data lie on the zero imaginary axis forming a line without spectral winding, except for a small portion (12\%) of points deviate from this axis (Fig.~\ref{fig:4}a top), attributed to experimental uncertainties, such as a low signal-to-noise ratio for quasi-momentum far from the center of the atom cloud. The observation is consistent with the simulation (Fig.~\ref{fig:4}a bottom). 

The results are in sharp contrast after opening up dissipation. We observe the nonzero spectral area and nontrivial spectral winding at a specific $k_x$ layer with dissipation $\gamma_\downarrow=0.4E_r$ (Fig. \ref{fig:4}b left), in agreement with simulated results obtained through plane-wave expansion (Fig.\ref{fig:4}b right). Notably, the spectral winding can be determined by measuring only the local complex band gap at each momentum in the experiment. The original band spectrum difference exhibits a ring-like pattern with an elliptical shape on the complex plane, and the sign of the imaginary part of the band gap changes as the quasi-momentum traverses the band inversion surface. By flipping the portion with a negative imaginary band gap into the first quadrant, the elliptical pattern evolves into the butterfly shape depicted in Fig.\ref{fig:4}b with a simulated result being consistent. Additionally, in the inset of Fig.\ref{fig:4}b, we present the band difference on the complex plane at $\gamma_\downarrow=0.8E_r$ showing the similar spectral winding. }

In Fig.~\ref{fig:4}c, we also present the experimental data and simulation at a particular $q_y$ layer while varying $k_x$ from $-0.6k_0$ to $k_0$. In this direction, the absence of optical lattice results in the band difference spectrum intersecting at $k_x\rightarrow\pm\infty$. Then, for a finite range of the band difference spectrum, the spectral flow is generally observable and can be measured in experiments, in sharp contrast to the case with lattice. 
Importantly, for specific $q_y$ layers ($q_y/k_0=-0.2$), the currently measured region is sufficient to allow the observation of the non-backstepping behavior, implying the existence of NHSE.


The spectral winding or non-backstepping behavior of the band difference $|\Re(\Delta E)|-|\Im(\Delta E)|$, as measured in the present experiment for quasi-1D layers, uniquely corresponds to that of the band dispersion. 
This guarantees that (i) the nontrivial spectral winding of the energy bands $E_{1/2}(k_x,q_y)$ and (ii) the nonzero spectral area (See Supplementary Information for a generic proof). Thus the measurement confirms the realization of the NSHE in both $x$ and $y$ directions, 
giving 2D corner skin modes under the OBC, which are actually supported by the numerical simulation in Fig.~\ref{fig:4}d. We note that in addition to the measured spectral winding and non-backstepping feature, 
the nonzero periodic-boundary spectral area [the dots in light color of all $k_x/k_0$ ($q_y/k_0$) in Fig.~\ref{fig:4}b (c)] also signals clearly the 2D skin effect in experiment.
 

\red{To further validate the presence of the NHSE in our system, we investigate its dynamical signature in real space by monitoring the spatial distribution of spin-up atoms (Fig. \ref{fig:5}a). Initially, we prepare the atoms with an initial velocity in the +x direction by quenching the system to a Hermitian Hamiltonian. Subsequently, the system is either shifted to a non-Hermitian Hamiltonian or maintained in the Hermitian Hamiltonian. After a certain hold time, we obtain the real-space profiles of the spin-up atoms by magnifying the {\it in-situ} image using a matter-wave magnifier~\cite{asteria2021quantum}, established by applying a harmonic trap along the z-direction.}

\red{In the first 1.5~ms, we observe the atom cloud moving in the +x direction for both cases. Following this, in the presence of the skin effect ($\gamma_\downarrow=0.2E_r$), the atoms gradually localize in the +x direction over time with slight movement, which is a direct consequence of the large number of skin modes accumulated in the +x boundary. Conversely, in the absence of the skin effect ($\gamma_\downarrow=0.0E_r$), the atoms move backward to their original position, exhibiting distinct dipole oscillation (Fig. \ref{fig:5}b,c). These observations are well consistent with the numerical simulation (Fig. \ref{fig:5}d,e), providing clear evidence for the manifestation of NHSE with nonzero dissipation.}

\red{To quantify the NHSE, we extract the center of mass from the 1D normalized spin-up atom profiles and further obtain the difference of center position between the Hermitian and non-Hermitian case (Fig. \ref{fig:5}f). With the skin effect, the center of mass position rises and then saturates. Without it, however, the position first rises and then falls, returning to its original position. In addition, we also take another experiment by initializing the motion along $-x$ direction. In the presence of skin effect in the +x boundary ($\gamma_\downarrow=0.2E_r$), the atom cloud stopped the motion to $-x$ direction in a relatively short time and then turned to $+x$ direction. In contrast, without the skin effect ($\gamma_\downarrow=0.0E_r$), the atom cloud can move to $-x$ direction with a larger displacement and exhibit dipole oscillation. The results show an effective bias force due to the skin effect to +x direction, and are also well consistent with the numerical results (See Supplementary Information for details). These observations further confirm the existence of skin effect with skin modes mainly accumulated to the +x direction in our present system. Notably, regardless of our setup only having a harmonic potential, the skin effect remains robust in external trapping potentials~\cite{guo2021theoretical}. It is suggested that systems with non-Hermitian spin-orbit coupling show a dynamic sticky effect in a box potential, resulting in unusual reflection behavior near the boundary~\cite{guo2021theoretical}.}
	
\red{Lastly, we have two remarks on the 2D NHSE in our system. One is that the occurrence of the NHSE could be inferred from the imaginary vector potential and imbalanced hoppings induced by SOC together with the dissipation (see Method). As shown by the density profile of eigenstates in the harmonic trap in Fig.~\ref{fig:4}d, besides the majority of skin modes being localized in the $(+x,+y)$ direction, some skin modes are localized in the $-x$ direction. In fact if the soft harmonic trap is replaced by a hard box trap, both $(\pm x,\pm y)$ corners will be populated with skin modes (see Supplementary Information). This feature originates from a universal result established by the particle-hole(-like) symmetry (PHS) that the skin modes related by the PHS are being localized in different boundaries~\cite{liu2023PHS}. Another thing worthwhile to mention is that, with appropriate approximations, the two low-energy sectors of our system could be mapped to Hermitian systems of free particles in 4D curved spaces that is the Cartesian products of two 2D pseudospheres (see Method). Together with the time dimension, the curved space after mapping is indeed (4+1)D. The nonzero curvature of the pseudospheres changes the inner product of states and renders the eigenstates to be localized in the boundaries of the curved spaces. This accumulation in the boundaries of the curved spaces is exactly the counterpart of the non-Hermitian skin effect in the mapping. The two dimensions of the 4D space can be considered as "extra dimensions" controlling the strength of the "geometric" potential experienced by the particles. This geometric description exhibits intriguing similarities with the (4+1)D Kaluza-Klein theory, a formalism unifying gravitation and electromagnetism through an extra dimension complementing to 3+1 space-time dimensions~\cite{Kaluza2018OnTU,Klein1926fiveD,Klein1926atom}. The potential of our non-Hermitian lattices shall inspire the further simulation of exotic high-dimensional curved spaces in future work.}

{\bf Conclusion}
{
We have observed for the first time 2D NHSE for an ultracold fermion gas, and investigated systematically the coexistence of the EPs and skin effect in the quantum system. Our realization is achieved by engineering the non-Hermiticity of the 2D Fermi gas with SOC, which shows to be an intriguing system to explore rich phenomena including the 2D topological nodes, skin effect, and EP physics. In addition to the static regime, the present realization opens up a promising direction to study the non-equilibrium quantum dynamics with engineered non-Hermiticity. In contrast to classical systems, our experiment naturally sets a quantum many-body system, and therefore paves a way to investigate the non-Hermitian quantum dynamics in the many-body regime using ultracold fermions with dissipation. Moreover, the high controllability also makes our system be a versatile platform to explore the high-dimensional non-Hermitian phenomena linking to profound physics like curved space~\cite{duality2022, Hermitization2022, zhou2022curve} and simulation of black holes~\cite{spinblackhole2012, Weylblackhole2022}, opening a broad avenue in studying exotic quantum physics beyond condensed matter and ultracold atoms. 
}


\clearpage
\newpage
\vspace{.1in} \noindent
\vspace{.05in} \\

{\bf 2D topological bands} In our experiments, the optical Raman lattice is generated by using a 1D spin-dependent optical lattice together with another free-running Raman beam to induce the Raman coupling, as described in Fig.\ref{fig:1}a. The lattice potential is produced by retroreflecting a Gaussian laser beam along y direction with the polarization linearly polarized along x direction, forming a spin-dependent potential $V_\sigma(y)=V_{0,\sigma}\cos^2{(k_0y)}$. The free-running linearly polarized Raman beam with polarization  direction z is applied along the direction with angle $\theta=76^o$ tilted from the lattice incident beam, inducing the Raman coupling $M_R(x,y)=M_0\cos{(k_0y)}e^{ik_2x-ik_1y}$ between $\ket{\uparrow}=\ket{m_F=5/2}$ and $\ket{\downarrow}=\ket{m_F=3/2}$, where $k_1=k_0\cos\theta$ and $k_2=k_0\sin\theta$. Both lattice beam and Raman beam are blue detuned by $\sim$1GHz from the $^1S_0(F=\frac{5}{2})\rightarrow{^3}P_1(F'=\frac{7}{2})$ transition of $^{173}\rm{Yb}$ atoms and the frequency difference $\delta\omega$ between the lattice lights and Raman lights can be tuned by acousto-optical modulators. The quantized axis is fixed by the bias magnetic field of 13.6G along the z-direction, which will induce an 8.1MHz Zeeman splitting between adjacent $m_F$ states. To separate out a spin-1/2 subspace from other hyperfine states of the ground manifold, we also use an additional $\sigma^-$ polarized 556 light along -z direction to induce spin-dependent ac stark shift(not shown in Fig.\ref{fig:1}b)\cite{song2016spin}. The spin dependent atom loss $\gamma_{\downarrow (\uparrow)}$ in our experiment is induced by a near-resonant $\sigma^-$ polarized loss beam with frequency red detuned by $\Delta_\downarrow=1.0\sim1.6(9.1\sim9.7)$MHz from $\ket{m_F=3/2}\rightarrow\ket{m_{F'}=1/2}$ transition ($\ket{m_F=5/2}\rightarrow\ket{m_{F'}=3/2}$ transition). Remarkably, the loss beam also induces an additional energy shift in two-photon detuning, which we can compensate by tuning the frequency difference between the Raman beam and the lattice beam in our experiment.

\red{To sum up, the s-band Bloch Hamiltonian could be written as:
\begin{align}    H_s(\mathbf{k})&=h_0(\mathbf{k})\sigma_0+h_z(\mathbf{k})\sigma_z+h_y(\mathbf{k})\sigma_y+i\gamma_\downarrow/4\sigma_z,\label{eq.sBloch}
\end{align}
where $\mathbf{k}=(k_x,q_y)$, $h_y(\mathbf{k})=2t_{so}\sin(q_y a_0-k_1a_0/2)$, $h_z(\mathbf{k})=-\hbar^2(k_x k_2)/2m-t_y^\uparrow\cos(q_y a_0)-t_y^\downarrow\cos(q_y a_0-k_1 a_0)+m_z$, $h_0(\mathbf{k})=\hbar^2(k_x^2+k_2^2/4)/2m-t_y^\uparrow\cos(q_y a_0)+t_y^\downarrow\cos(q_y a_0-k_1 a_0)-i\gamma_\downarrow/4$ and $\sigma_{0,y,z}$ are the Pauli matrices. $t_y^{\uparrow(\downarrow)}$ is the nearest hopping coefficient of spin-up(down) atoms and $a_0=\pi/k_0$ is the lattice constant in $y$ direction.}

\textbf{Experimental procedure} We start the experiment with a single-component degenerate $^{173}$Yb Fermi gas at $T/T_F\lesssim0.4$ with atom number around $2\times10^4$ in a far detuned crossed optical dipole trap with trap frequencies $\bar{\omega}=(\omega_x\omega_y\omega_z)^{1/3}=112\times2\pi$Hz.
The single-component degenerate $^{173}\text{Yb}$ Fermi gas in the $\ket{\uparrow}$ state is prepared with the optical pumping process during and after the evaporation cooling. The atoms are then loaded into our 1D optical lattice potential by ramping up both the lattice and Raman laser intensity in 3ms with the initial two-photon detuning $\delta_i$ far enough to exclude the Raman coupling. After that, we further hold the system for another 1ms and then quench the two-photon detuning from $\delta_i$ to $\delta_f$ at time $t=0$ and switch the loss pulse at the same time. Meanwhile, the optical dipole trap is switched off in case it affects the evolution. The atoms initially prepared in $\ket{\uparrow}$ state start the Raman-Rabi oscillation between the lowest two bands in our system. 

Holding a variable time $t_f$, we record spin-sensitive momentum distribution after 10ms (or 15ms) time of flight expansion by applying the 556nm blast pulses resonant to $^1S_0(F'=5/2)\rightarrow{}^3P_1(F'=7/2)$ transition. Three absorption images with no atoms killed $\mathcal{I}_1$, with spin up atoms killed $\mathcal{I}_2$ and with both spin up and spin down atoms killed $\mathcal{I}_3$ are obtained and we can further extract the atomic distribution of spin up and spin down from $\mathcal{I}_\uparrow(k'_x,k_y)=\mathcal{I}_1-\mathcal{I}_2$ and $\mathcal{I}_\downarrow(k'_x,k_y)=\mathcal{I}_2-\mathcal{I}_3$. After folding the atomic distribution of two spin states into the first Brillouin zone and shifting the momentum distribution of the spin-down state based on the spin momentum locking, we obtain the momentum distribution in quasi-momentum $\mathcal{D}_{\sigma}(k_x,q_y)$. Finally, the spin polarization can be calculated by $P(k_x,q_y)=\frac{\mathcal{D}_\uparrow(k_x,q_y)-\mathcal{D}_\downarrow(k_x,q_y)}{\mathcal{D}_\uparrow(k_x,q_y)+\mathcal{D}_\downarrow(k_x,q_y)}$.

{\bf Fitting procedure} When the system is Hermitian, the spin polarization oscillates between the lowest two dressed states following Rabi process which can be described by a sinusoidal function. As the loss rate increases, we find that the time evolution gradually change to a monotonic exponential decay curve (Fig.~\ref{fig:3}c). To obtain the Phase diagram of the PT-symmetry breaking and further explore the shift of EPs at different quasi-momentum, one can extract the energy band gap $\Re{(\Delta E)}$ and damping rate $\Im{(\Delta E)}$ with a damped sinusoidal fitting function. However, when the dissipation is large, the damping term dominates the time evolution curve and the fitting is not sensitive to the oscillation term. Therefore, we quantify the different behaviors before and after PT-symmetry breaking transition by fitting with both damped oscillation and exponential decay and compare the $R^2=1-\frac{\sum_{i=1}^{n}(y_i-f(x_i))^2}{\sum_{i=1}^{n}(y_i-\bar{y})^2}$. When the loss rate is large enough, the $R^2$ ratio between these two fits is close to 1, which indicates the unnecessary of the oscillation term when fitting the time evolution at a large loss (Fig.~\ref{fig:3}b). With this method, we can estimate where the transition occurs and then fit the time evolution curve before and after the transition region with sinusoidal function and exponential function respectively to extract the energy band and damping rate (Fig.~\ref{fig:3}a). For different $k_x/k_0$ (or equivalently $\delta$), the loss rates of EPs are different.

\textbf{Reconstruction of complex eigen spectrum} The momentum uncertainty caused by the finite resolution of our imaging system and fluctuation of two-photon detuning will make the oscillation of polarization dephase even without spin-dependent dissipation. At the low dissipation regime, fitting the oscillation of spin polarization can not distinguish this type of dephasing from the decay induced by spin-dependent dissipation. To avoid this issue, when we reconstruct the complex eigenspectrum in the low dissipation regime, the imaginary part of the energy gap is determined by fitting the oscillation curve of spin-down atoms with the following equation
\red{
\begin{equation}
\begin{aligned}
f(t)&=e^{2\Im\left(E_1\right)t}\Big[ A(1+e^{2\Im\left(E_2-E_1\right)t})\\
&-{2Be}^{\Im\left({E_2-E_1}\right)t}\cos{\Re\left(E_2-E_1\right)t} \Big]\\
&+ 2B-2A
\end{aligned}
\end{equation}}

\noindent\red{ in which $\Im(E_2-E_1)$ is the imaginary band gap and the dephasing induced by the momentum uncertainty can be absorbed by the $e^{2\Im(E_1)t}$ term without influencing the imaginary band gap in the low dissipation regime. During the fitting process, the initial values of the fitting parameters are determined from the damped oscillation fitting of the corresponding spin polarization data. This aids in guiding the fitting towards a solution that aligns with the appropriate physical parameters of our system. Further combined with the real part extracted from the oscillation curve of spin polarization, the complex eigenspectrum can be reconstructed.}

\textbf{Numerical simulation in the real space} 
To see the NHSE of our optical lattice system, here we solve a 2D lattice model with open boundary cut along $x$ and $y$ directions. The model is obtained by applying s-band tight-binding approximation in the $y$ direction and finite-difference approximation in the $x$ direction on the experimental model in Eq.~\eqref{expmodel}. It acts on the local basis $|\mathbf{j}s\rangle$ centered at the site $\mathbf{j}=(j_x,j_y)$ with spin $s=(\uparrow,\downarrow)$, which in the $y$ direction are the s-band Wannier functions. The Hamiltonian is
    \red{\begin{align}
        H_{L}=&\sum_{\mathbf{j}}\Big\{m_z\ket{\mathbf{j}\uparrow}\bra{\mathbf{j}\uparrow}-(m_z+i\gamma_\downarrow/2)\ket{\mathbf{j}\downarrow}\bra{\mathbf{j}\downarrow}\nonumber\\      
        &-\sum_{k=x,y}\Big[\bigr(t_{\uparrow}^k\ket{\mathbf{j}+\Vec{e}_k\uparrow}\bra{\mathbf{j}\uparrow}-t_\downarrow^k\ket{\mathbf{j}+\Vec{e}_k\downarrow}\bra{\mathbf{j}\downarrow}\bigr)\nonumber\\
        &+h.c.\Big]+\bigr(t_{so}e^{-ik_0 \cos\theta j_y}\ket{\mathbf{j}\downarrow}\bra{\mathbf{j}+\Vec{e}_y\uparrow}\nonumber\\
        &-t_{so}e^{ik_0 \cos\theta j_y}\ket{\mathbf{j}\uparrow}\bra{\mathbf{j}+\Vec{e}_y\downarrow}+h.c.\bigr)+V_{tr}(\mathbf{j})\Big\},\label{latticemodel}
    \end{align}}
    where $m_z$ is the Zeeman energy, $\gamma_\downarrow$ is the loss rate of spin-down atoms, $t_{\uparrow,\downarrow}^k$ are the nearest-neighbor spin-conserved hopping coefficients of the optical lattice in the $k=(x,y)$ directions, $t_{so}$ is the nearest-neighbor spin-flipped hopping coefficient in the optical lattice direction, and $\theta=76^o, k_0=\pi/a_0$ with $a_0$ being the lattice constant in the $y$ direction. \red{Here $t_{\uparrow}^x={t_{\downarrow}^x}^*=i \sin\theta \frac{a}{2l_x \pi}+\frac{a^2}{l_x^2 \pi^2}$ where $l_x$ is the finite difference step-length. We could gauge out the phase factor $e^{\pm i k_0 \cos\theta j_y}$ with a transformation $\ket{\mathbf{j}\downarrow}\rightarrow e^{i k_0 \cos\theta j_y}\ket{\mathbf{j}\downarrow}$ and obtain a commensurate lattice system explicitly. 
    $V_{tr}(\mathbf{j})=\frac{m\omega_x^2j_x^2}{2l_x^2}+\frac{m\omega_y^2j_y^2}{2a^2}$ is the trap potential.} This 2D lattice model approximates the low-energy and long wavelength sector of the original system well, but may fail in the high-energy and short wavelength parts. Fig.~\ref{fig:4}(d) shows the NHSE in the real space, in which the tight-binding parameters are set to be $t_{\uparrow(\downarrow)}=0.32(0.38)E_r$ and $t_{so}=0.25E_r$ corresponding to the experimental parameters $V_{\uparrow(\downarrow)}=3.2(2.2)E_r$ and $M_R=2.0E_r$. The dissipation rate is $\gamma_\downarrow=0.2E_r$ and the Zeeman splitting is $m_z=-0.3E_r$. The step-length in the $x$ direction is chosen to be $x_0=a_0/7$ and we see that the low-energy solutions within the s-band regime of the optical lattice have converged well enough (See Supplementary Information).

\red{\textbf{Spatial distribution signal detection} To observe the signal of the NHSE in real space, we employ a quantum gas magnifier in our system~\cite{asteria2021quantum}. This magnifier is created by applying a far-detuned optical dipole trap along the z-direction prior to the absorption imaging process, which effectively establishes a harmonic trap in the x-y plane. By applying a harmonic potential with a trapping frequency of $\omega_{\text{M}} = 2\pi/T$ for a duration of $T/4$, we map the spatial distribution to the momentum distribution~\cite{murthy2014matter, holten2022observation}. Combining this with a subsequent free time-of-flight (TOF) expansion, we reproduce the initial spatial distribution with a magnification factor of approximately $M\approx\omega_{\text{M}}t_{\text{TOF}}$. For our specific experimental parameters, with $t_{\text{TOF}}=10\text{ms}$ and $\omega_{\text{M}}/(2\pi)=181\text{Hz}$, the quantum gas magnifier provides an additional magnification of approximately 11.4 on top of the magnification achieved through conventional absorption imaging with a magnification of 3.73, with the estimated resolution along 0.2$\mu m$. It is important to note that the current magnification is not sufficient to resolve the lattice spacing. Therefore, during the calibration of the focusing condition, we hold the atom cloud in the harmonic trap for a time of $T/2$, resulting in a momentum distribution opposite in direction to the initial momentum distribution.}

 \red{Utilizing the quantum gas magnifier, we proceed to investigate the manifestation of NHSE in real space. It is important to note that when the atoms are located far away from the boundary, the presence or absence of the skin effect may have limited influence on the short-term evolution of the atom cloud~\cite{guo2021theoretical, mao2021boundary}. Therefore, we initially quench the atoms in the crossed dipole trap to the Hermitian optical Raman lattice and allow them to evolve for a specific duration, establishing an initial state with a velocity in the +x direction. Subsequently, the system is either subjected to another quench to the non-Hermitian Hamiltonian or maintained in the original Hermitian Hamiltonian. After a certain hold time, we perform real-space detection of the spin-up atoms using the quantum gas magnifier. To mitigate perturbations arising from misalignment between the magnifier beam and the cross optical dipole trap during the ramping of the magnifier, the magnifier beam is ramped up prior to preparing the initial state.}

\red{\textbf{Mapping to 4D curved space} Here we consider the low-energy models of our experimental system and establish the mapping to Hermitian models of free particles living in 4D curved spaces. We start from the s-band Bloch Hamiltonian in Eq.~\eqref{eq.sBloch}. For simplicity, we assume $t_{\uparrow}^y=t_{\downarrow}^y=t_y$ and apply an overall $(k_0+k_1)/2$ quasi-momentum shift in the $y$ direction. The coefficients are simplified as 
$h_y(\mathbf{k})=-2t_{so}\cos(q_y a_0)$, $h_z(\mathbf{k})=-\hbar^2(k_xk_2)/2m-2t_y\sin(q_y a_0)\cos(k_1 a_0/2)+m_z$, $h_0(\mathbf{k})=\hbar^2(k_x^2+k_2^2/4)/2m-2t_y\cos(q_y a_0)\sin(k_1 a_0/2)-i\gamma_\downarrow/4$. To explicitly reveal the origin of the non-Hermitian skin effect and the connection to curved spaces, we rewrite the model under the spin-orbital-coupled Bloch bands $|\pm,\mathbf{k}\rangle$, which diagonalize the Hermitian part of the Hamiltonian as $[H(\mathbf{k})-i\gamma_\downarrow/4\sigma_z]|\pm,\mathbf{k}\rangle=[h_0(\mathbf{k})\pm \epsilon(\mathbf{k})]|\pm,\mathbf{k}\rangle$ with $\epsilon(\mathbf{k})=\sqrt{h_z^2+h_y^2}$. The model is rewritten as
     \begin{align}
        \tilde{H}(\mathbf{k})&=h_0(\mathbf{k})+\epsilon(\mathbf{k})\sigma_z+i\frac{\gamma_\downarrow}{4}\frac{h_z(\mathbf{k})\sigma_z+h_y(\mathbf{k})\sigma_y}{\epsilon(\mathbf{k})}.\label{eq.newmodel}
    \end{align}
    The $i\frac{\gamma_\downarrow h_z(\mathbf{k})}{4\epsilon(\mathbf{k})}\sigma_z$ term contributes imaginary vector potential in the $x$ direction and imbalanced hopppings in the $y$ direction, both of which lead to NHSE. We restrict the following discussion in the long-wavelength and weak dissipation regime, i.e. $|k_x|,|q_y|\rightarrow 0$ and $\gamma_\downarrow<|m_z|$. Then $\epsilon(\mathbf{k})\approx \sqrt{m_z^2+4t_{so}^2}:=E_0$ and the off-diagonal term in Eq.~\eqref{eq.newmodel} could be neglected (see Supplementary Information for details). Then the Hamiltonian is diagonalized by $\ket{\pm,\mathbf{k}}$, with diagonal terms
    \begin{align}
        \tilde{H}_\pm(x,y)&=\hbar^2(-i\partial_x\mp iA_x)^2/(2m)\nonumber\\
        &+\hbar^2(-i\partial_y\mp iA_y)^2/(2m_y)\pm(E_0+\frac{\gamma_\downarrow \delta_z}{4E_0}),\label{eq.Hpm}
    \end{align}
    where $(A_x,A_y)=(\frac{\gamma_\downarrow k_2}{8E_0},\frac{\gamma_\downarrow \cot(k_1a_0/2)}{4a_0 E_0})$ and $m_y=2 m t_y\sin(k_1 a_0/2)a_0^2/\hbar^2$ is the effective mass at $q_y=0$. We notice that this could be mapped to Hermitian models on curved spaces. Consider the 4D curved spaces $M^\pm(\zeta,\eta,\zeta',\eta')$ with metrics
    \begin{align}
    \mathbf{g}_\pm(\zeta,\eta,\zeta',\eta')=&(\zeta^{-2}d\zeta^2+\zeta^{\mp2}d\zeta'^2)/(2A_x)^2\nonumber\\
    &+(\eta^{-2}d\eta^2+\eta^{\mp2}d\eta'^2)m_y/(4mA_y^2),
    \end{align} respectively. The Hamiltonian of free particles are given by 
    \begin{align}
        H_\text{cv}^\pm(\zeta,\eta,\zeta',\eta')=&-\frac{\hbar^2}{2m}\frac{1}{\sqrt{g_\pm}}\sum_{j,k}\partial_j\mathbf{g}^{jk}_\pm\sqrt{g_\pm}\partial_k,
    \end{align}
    with $j,k$ running over $(\zeta,\eta,\zeta',\eta')$ and $g_\pm=\det(\mathbf{g_\pm})$. With a coordinate transformation $(\zeta,\eta)=(e^{-2A_x x},e^{-2A_y y})$, the low-energy Hamiltonians in Eq.~\eqref{eq.Hpm} are mapped to Hermitian systems of the free particles on curved spaces, namely we have
    \begin{equation}
        \tilde{H}_\pm(x,y)=H^\pm_{\text{cv}}(\zeta,\eta,k_{\zeta'}=0,k_{\eta'}=0),
    \end{equation}
    where $k_{\zeta',\eta'}$ are the momenta components in $\zeta',\eta'$ directions. $M^\pm(\zeta,\eta,\zeta',\eta')$ are the Cartesian products of two pseudospheres. The eigenstates of $H^\pm_{\text{cv}}(\zeta,\eta,k_{\zeta'}=0,k_{\eta'}=0)$ are localized on the corners of $(\zeta,\eta)$ under OBC, in agreement with the corner skin effect of $\tilde{H}_\pm$.}

\vspace{.1in} \noindent
\textbf{ACKNOWLEDGMENTS}
GBJ acknowledges support from the RGC through 16306119, 16302420, 16302821, 16306321, 16306922, C6009-20G, N-HKUST636-22, and RFS2122-6S04. XJL was supported by National Key Research and Development Program of China (2021YFA1400900), the National Natural Science Foundation of China (Grants No. 11825401 and No. 12261160368), the Innovation Program for Quantum Science and Technology (Grant No. 2021ZD0302000).
\vspace{.05in} \\

\pagebreak
\newpage
\clearpage
\onecolumngrid

\setcounter{section}{0}  
\setcounter{equation}{0}  
\setcounter{figure}{0}  

\renewcommand{\figurename}{Figure S}
\renewcommand\refname{Supplementary References}

\title{SUPPLEMENTARY INFORMATION\\
	for\\
	"Two-dimensional non-Hermitian skin effect in an ultracold Fermi gas"}

\author{Entong Zhao}
\thanks{These authors contributed equally to this work.}
\affiliation{Department of Physics, The Hong Kong University of Science and Technology,\\ Clear Water Bay, Kowloon, Hong Kong, China}

\author{Zhiyuan Wang}
\thanks{These authors contributed equally to this work.}
\affiliation{International Center for Quantum Materials, School of Physics, Peking University, Beijing 100871, China}

\author{Chengdong He}
\affiliation{Department of Physics, The Hong Kong University of Science and Technology,\\ Clear Water Bay, Kowloon, Hong Kong, China}

\author{Ting Fung Jeffrey Poon}
\affiliation{International Center for Quantum Materials, School of Physics, Peking University, Beijing 100871, China}

\author{Ka Kwan Pak}
\affiliation{Department of Physics, The Hong Kong University of Science and Technology,\\ Clear Water Bay, Kowloon, Hong Kong, China}

\author{Yu-Jun Liu}
\affiliation{Department of Physics, The Hong Kong University of Science and Technology,\\ Clear Water Bay, Kowloon, Hong Kong, China}

\author{Peng Ren}
\affiliation{Department of Physics, The Hong Kong University of Science and Technology,\\ Clear Water Bay, Kowloon, Hong Kong, China}

\author{Xiong-Jun Liu}
\email{xiongjunliu@pku.edu.cn}
\affiliation{International Center for Quantum Materials, School of Physics, Peking University, Beijing 100871, China}
\affiliation{Hefei National Laboratory, Hefei 230088, China}
\affiliation{International Quantum Academy, Shenzhen 518048, China}

\author{Gyu-Boong Jo}
\email{gbjo@ust.hk}
\affiliation{Department of Physics, The Hong Kong University of Science and Technology,\\ Clear Water Bay, Kowloon, Hong Kong, China}
\affiliation{IAS Center for Quantum Technologies, The Hong Kong University of Science and Technology, Hong Kong, China}

\maketitle

\begin{widetext}
	\section{Experimental setup and procedure}
	\subsection{Experimental setup}

	The spin-dependent 1D optical lattice potential along $y$ direction is generated by retro-reflecting a Gaussian laser beam with the frequency $\omega$ blue detuned by 1GHz from the $^1S_0(F=\frac{5}{2})\rightarrow{^3}P_1(F'=\frac{7}{2})$ transition of $^{173}\rm{Yb}$ atoms and the polarization linearly polarized along $x$ direction (Figure S\ref{FigS0}). The electromagnetic field of the lattice beam can be written as $\mathbf{E_y}\propto\hat{e}_xE_y\cos(k_0y+\omega t+\phi)=\mathbf{E_++E_-}$, with $\mathbf{E}_{\pm}=\frac{1}{\sqrt{2}}\hat{e}_\pm E_y\cos(k_0y+\omega t+\phi)$ and $\hat{e}_\pm=\frac{1}{\sqrt{2}}(\hat{e}_x\pm i\hat{e}_y)$, forming the lattice potential 
	\begin{eqnarray}
		V^{\text{latt}}_{\sigma}(y)=\sum_{F'}\left(\frac{\hbar\Omega^2_{y,\sigma,+,F'}}{4\Delta_{y,F'}}+\frac{\hbar\Omega^2_{y,\sigma,-,F'}}{4\Delta_{y,F'}}\right)=V_{\sigma}\cos^2(k_0y),
	\end{eqnarray}
	where the effective Rabi frequency $\Omega_{y,\sigma,\pm,F'}=\langle F,\sigma|e\frac{x+y}{\sqrt{2}}|F',m_F\pm1\rangle E_{\pm}$ and single-photon detunings $\Delta_{y,F'}$ are determined from all relevant transitions to the excited states $(F'=\frac{7}{2},\frac{5}{2},\frac{3}{2})$ in the $^3P_1$ manifold. 
	
	Another linearly ($\pi$) polarized light with polarization direction $z$ and frequency $\omega'=\omega+\delta\omega$, denoted as $\mathbf{E_x}\propto\hat{e}_zE_xe^{-ik_1y+ik_2x+\omega' t+\phi'}$, is applied along the direction with angle $76^o$ tilted from the lattice incident beam and induces the Raman coupling between $\ket{\uparrow}=|m_F=\frac{5}{2}\rangle$ and $\ket{\downarrow}=|m_F=\frac{3}{2}\rangle$, where $k_1=k_0\cos\theta$ and $k_2=k_0\sin\theta$.  The frequency difference $\delta\omega$ between the lattice lights and Raman light can be tuned by the acousto-optical modulators. The Raman coupling potential $M_R$ has the form 
	\begin{eqnarray}
		M_{R}(x,y)=\sum_{F'}\frac{\hbar\Omega_{x,\sigma,+,F'}\Omega_{R,\sigma,F'}}{4\Delta_{y,F'}}=M_0\cos(k_0y)e^{-ik_1y+ik_2x},
	\end{eqnarray}
	with effective Rabi frequency $\Omega_{R,\sigma,F'}=\langle F,\sigma|ez|F',m_F\rangle E_x$.
	
	\begin{figure}[htbp]
		\centering
		\includegraphics[width=0.65\columnwidth]{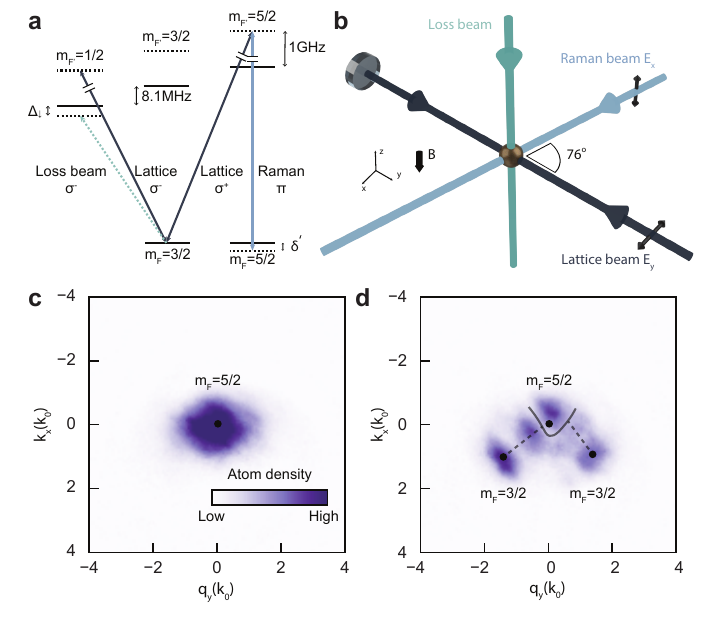}
		\caption{\textbf{a},  Schematic energy level diagram with relevant transitions. Both the lattice and the Raman beams are blue-detuned from $^1S_0(F=\frac{5}{2})\rightarrow{^3}P_1(F'=\frac{7}{2})$ intercombination transition and induce the Raman transition between the $\ket{\uparrow}=\ket{m_F=5/2}$ and $\ket{\downarrow}=\ket{m_F=3/2}$ hyperfine states of the $ ^1S_0$ ground manifold. \textbf{b}, Our optical Raman lattice system, which consists of a 1D optical lattice along the y direction and another Raman beam along the direction with angle $76^o$ tilted from the lattice incident beam, generating a periodic Raman potential. A loss beam with $\sigma^-$ polarization is further applied along -z direction to realize the non-Hermiticity. \textbf{c}, Absorption image of $m_F$ = 5/2 atoms without quenching the two photon detuning ($t_f$=0ms) after 15ms time-of-flight. \textbf{d}, When the two photon detuning is quenched from $\delta_i$ to $\delta_f$, two $m_F$ = 3/2 atom clouds are coupled out from $m_F$ = 5/2 clouds after holding $t_f$=0.15ms, which shows the resonant spin flipping on particular quasi-momentum subspace.}\label{FigS0}
	\end{figure}
	
	To separate out spin-1/2 space from other hyperfine states of the ground manifold, we use an additional $\sigma^-$ polarized 556nm light along the z direction called lift beam to induce spin-dependent ac stark shift. Notably, the spin-dependent lattice potential $V_{\uparrow(\downarrow)}=3.2{E_r}(2.2{E_r})$ will induce an additional on-site energy difference $\delta_0$ to the Bloch bands of two spin states, which can be compensated by the two-photon detuning. The quantized axis is fixed by the bias magnetic field of 13.6G along the z-direction, which will induce an 8.1MHz Zeeman splitting between adjacent $m_F$ states. The loss beam is $\sigma^-$ polarized and red detuned by $\Delta_\downarrow=1.0$ or 1.6MHz and $\Delta_\uparrow=9.1$ or 9.7MHz with respect to $^1S_0(F=\frac{5}{2},m_F=\frac{3}{2})\rightarrow{^3P_1}(F'=\frac{7}{2},m_{F'}=\frac{1}{2})$ and $^1S_0(F=\frac{5}{2},m_F=\frac{5}{2})\rightarrow{^3P_1}(F'=\frac{7}{2},m_{F'}=\frac{3}{2})$. Based on CG-coefficients, the estimated ratio of loss rate between two spin states $\gamma_\downarrow/\gamma_\uparrow$ is around 110 for $\Delta_\downarrow=-1.0$MHz and 248 for $\Delta_\downarrow=-1.6$MHz, which means the $\gamma_\uparrow$ term can be ignored. Remarkably, the loss beam will also induce an additional energy shift in two photon detuning, which we can compensate by tuning the frequency difference between Raman beam and lattice beam in our experiment.
	
	\begin{figure}[htbp]
		\centering
		\includegraphics[width=0.9\columnwidth]{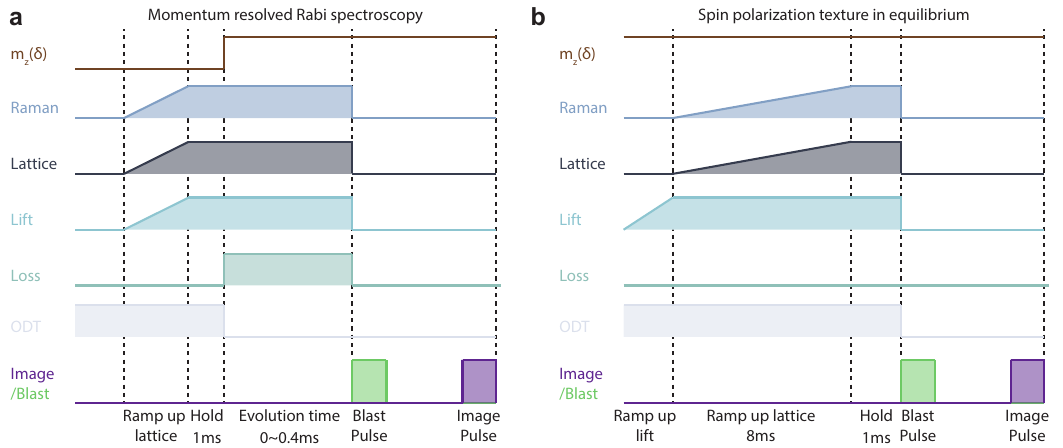}
		\caption{Experimental sequence for momentum resolved Rabi spectroscopy (\textbf{a}) and spin polarization texture measurement in equilibrium (\textbf{b}). }\label{FigS1}
	\end{figure}
	
	\subsection{Experimental procedure}
	
	The experimental sequence for the momentum resolved Rabi spectroscopy can be briefly described in Figure S\ref{FigS1}a. A single-component degenerate $^{173}\text{Yb}$ Fermi gas in the $\ket{\uparrow}$ state is prepared with the optical pumping process during (with 399nm light) and after (with 556nm light) the evaporation. The atoms are then loaded into a 1D optical lattice potential by ramping up both the lattice and Raman laser intensity in 3ms with the initial two photon detuning $\delta_i=-27E_r$ far enough to exclude the Raman coupling. The final lattice depth and Raman coupling strength are set to $V_{\uparrow(\downarrow)}=3.2(2.2)E_r$ and $M_0=2.0E_r$ ($\text{Er}=\hbar^2k_0^2/2m=h\times3.735\text{kHz}$), which can be calibrated from Kapitza-Dirac diffraction and the two-photon Rabi oscillation, respectively.  After the initial state preparation, we further hold the system for another 1ms and then quench the two-photon detuning from $\delta_i$ to $\delta_f$ at time $t=0$ by controlling the RF signal of the acousto-optical modulator and the loss pulse is switched on at the same time. The atoms initially prepared in $\ket{\uparrow}$ state start the Raman-Rabi oscillation between the lowest two bands in our system. Holding a variable time $t_f$, we record spin-sensitive momentum distribution after 10ms (or 15ms) time of flight expansion by applying the 556nm blast pulses resonant to $^1S_0(F'=5/2)\rightarrow{}^3P_1(F'=7/2)$ transition.
	
	In our Rabi spectroscopy, we found that there is no obvious difference between the lattice loading in 3ms and 9ms, probably because the final lattice depth is not quite deep. From the estimation of Kapitza-Dirac diffraction, even though the lattice is suddenly switched on, the maximum population for higher order is smaller than 30\% and the oscillation period is around $60\mu s$, much smaller than the hold time (1ms). Therefore, the shorter loading time will not affect the Rabi spectroscopy and will maximumly suppresses the heating and other detrimental effects. Besides, it is worth noting that the optical dipole trap is switched off at the same time as the quench dynamics starts. This is because, in previous experiments\cite{song2018ob,ren2022ch}, we found the intraband transition induced by optical dipole trap will affect the time evolution. Since the time evolution is smaller than 0.4ms, the atoms fall less than 1$\mu m$ under gravity, which is much smaller than the beam waist of the lattice and Raman beam and can be ignored in our experiment.
	
	\begin{figure}[htbp]
		\centering
		\includegraphics[width=0.9\columnwidth]{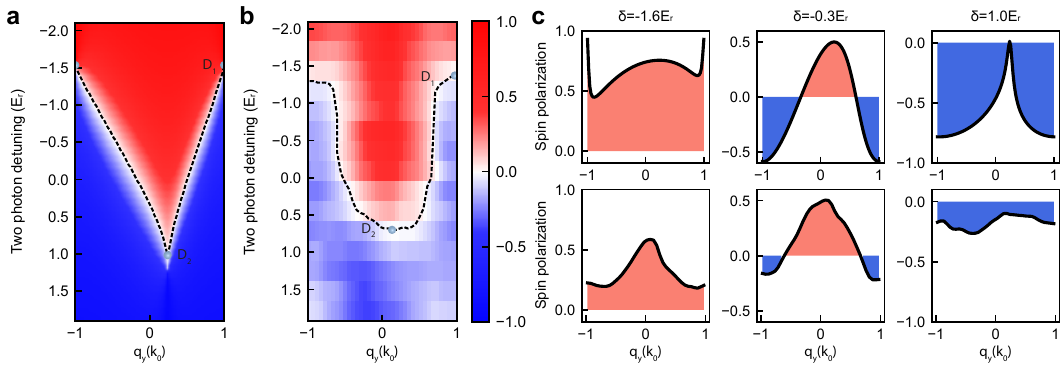}
		\caption{Spin polarization of equilibrium state. \textbf{a} Numerical simulation of spin polarization as a function of two photon detuning $\delta$. \textbf{b} Experimental measurement of spin polarization as a function of two photon detuning $\delta$. \textbf{c} Theoretical and experimental spin polarization texture of the lowest band for $\delta=-1.6, -0.3, 1.0E_r$, respectively. }\label{FigS2}
	\end{figure}
	
	\subsection{Spin polarization of equilibrium state in Hermitian regime}
	
	Except for the momentum-resolved Rabi spectroscopy, we also measured the spin polarization texture in equilibrium with experimental sequence in Figure S\ref{FigS1}b. An optical ac Stark shift is added by a lift beam 2ms before the optical lattice potential, which separates out an effective spin-1/2 subspace from other hyperfine levels. The Raman beam and lattice beams are adiabatically switched on with an 8-ms exponential ramp to the final value. The 10ms spin-resolved time-of-light image is taken after a spin sensitive blast sequence to reconstruct the spin polarization texture. Dependent on the value of two photon detuning (equivalent to $k_x$), our system can form either a topological band or trivial band along $q_y$ direction, as shown in Figure S\ref{FigS2}a. The experimentally measured spin polarization $P(q_y)$ can be reconstructed from the integrated momentum distribution along $k_x$ direction within $k_x/k_0=-0.2$ to $0.2$, as shown in Figure S\ref{FigS2}b, similar to the detection technique in \cite{song2019ob}. When the two photon detuning is in the trivial regime, the spin polarization is either $\ket{\uparrow}$(red)-dominated or $\ket{\downarrow}$(blue)-dominated. However, when the Raman transition resonantly couples the lowest two bands, spin polarization changes from blue to red and blue again within the first Brillouin zone, which indicates the nonzero winding number and topological band\cite{song2018ob}.(Figure S\ref{FigS2}c) Notably, since the atom number at higher $k_x$ is not enough for a good signal-to-noise ratio, we choose to fix $k_x$ and scan the value of two photon detuning instead of fixing the two photon detuning and changing $k_x$ to obtain the whole spin polarization texture in equilibrium case.
	
	\begin{figure}[htbp]
		\centering
		\includegraphics[width=0.9\columnwidth]{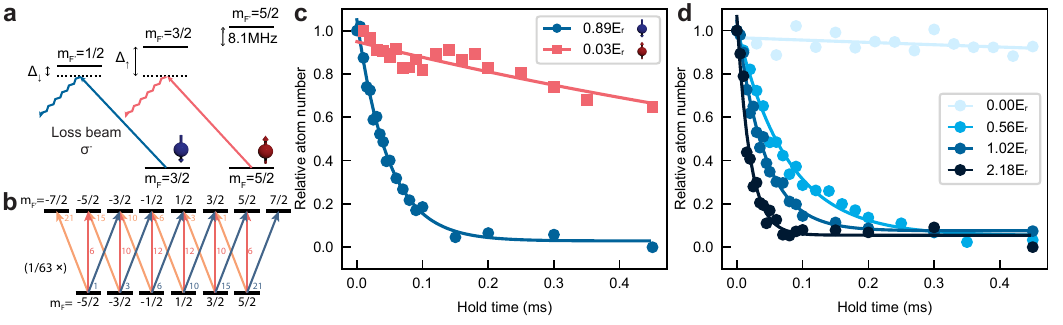}
		\caption{Calibration of the atomic loss. \textbf{a} Schematic energy level for spin dependent loss. \textbf{b} Relative optical transition strength of different $m_F$ states for the narrow intercombination transition $^1S_0(F=\frac{5}{2})\rightarrow{}^3P_1(F=\frac{7}{2})$. \textbf{c} Loss rate calibration for spin-down and spin-up state at same optical intensity. \textbf{d} Loss rate calibration for spin-down state at variable intensity of the loss beam.}\label{FigS3}
	\end{figure}
	
	\subsection{Calibration of the atomic loss}
	In our system, the spin dependent loss is induced by an optical pumping beam at a small detuning from the narrow intercombination transition $^1S_0(F=\frac{5}{2})\rightarrow{}^3P_1(F=\frac{7}{2})$, as shown in Figure S\ref{FigS3}a. The optical pumping beam is red detuned by $\Delta_\downarrow=1.0$ or 1.6MHz and $\Delta_\uparrow=9.1$ or 9.7MHz with respect to $\ket{\downarrow}\rightarrow\ket{m_{F'}=\frac{1}{2}}$ and $\ket{\uparrow}\rightarrow\ket{m_{F'}=\frac{3}{2}}$, together with the relative optical transition strength or CG-coefficients (Figure S\ref{FigS3}b), resulting in the spin dependence of atomic loss. In our experiment, we first prepare the atoms on specific spin state $\ket{\sigma}=\ket{\uparrow}$ or $\ket{\downarrow}$ in optical lattice and then suddenly switching on the loss pulse. The two photon detuning $m_z(\delta)$ is kept constant and large enough ($\delta\sim-134E_r$) to avoid the Raman coupling during the pulse, and  the relative atom number on specific state can then be resolved by optical Stern Gerlach measurment. The experimental sequence for other optical beams are same as that in momentum resolved Rabi spectroscopy measurement to avoid the effect of other types of dissipation. We further calibrate the loss rate by fitting the relative atom number of spin states $\ket{\sigma=\uparrow,\downarrow}$ with $f(t)=Ae^{-\frac{\gamma_\sigma}{\hbar} t}+B$ and the loss rate is tunable by controlling the power of the loss beam (Figure S\ref{FigS3}c). From the calibration, the loss rate between two spin states $\gamma_\downarrow/\gamma_\uparrow$ is larger than 20 for $\Delta_\downarrow=-1.6$MHz.
	
	\begin{figure}[htbp]
		\centering
		\includegraphics[width=0.9\columnwidth]{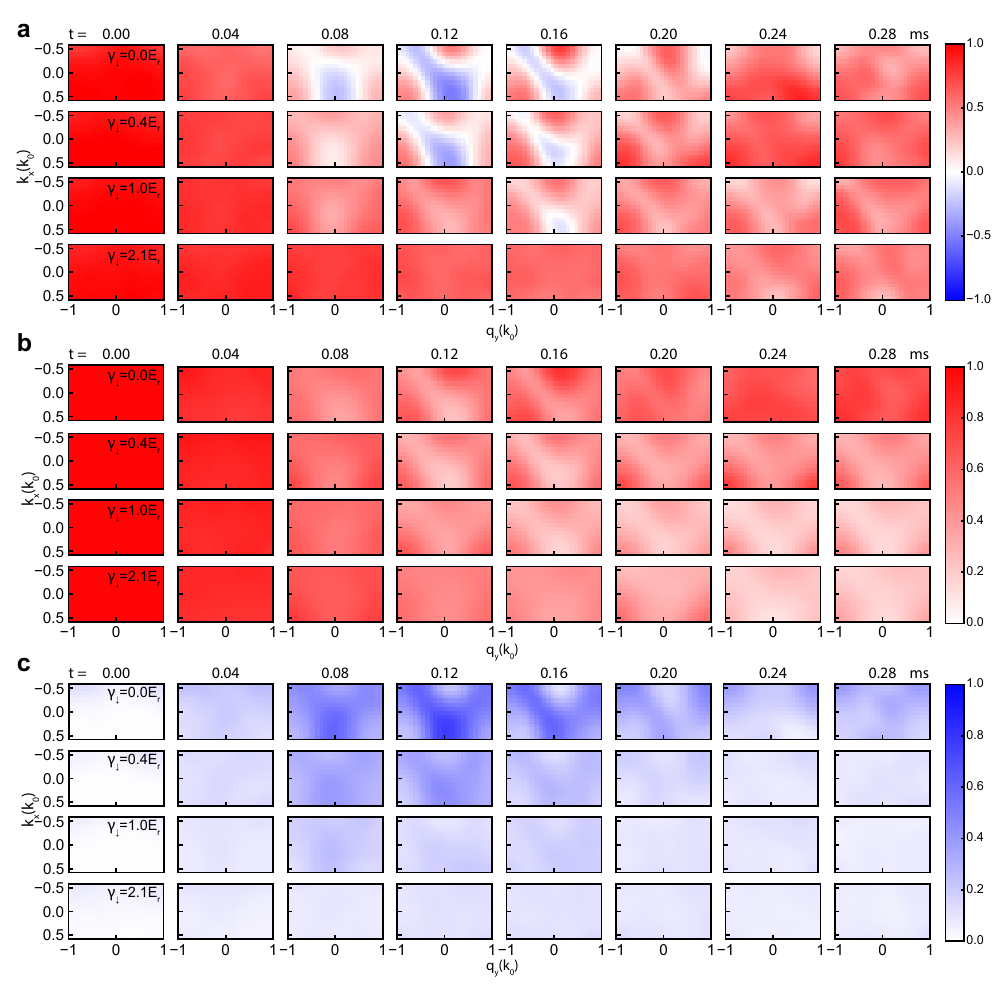}
		\caption{2D momentum dependent time evolution for spin polarization texture $P(k_x,q_y,t)$ in \textbf{a}, normalized spin up texture $\mathcal{D}_\uparrow(k_x,q_y,t)/\mathcal{D}_\uparrow(k_x,q_y,t=0)$ in \textbf{b}, normalized spin down texture $\mathcal{D}_\downarrow(k_x,q_y,t)/\mathcal{D}_\uparrow(k_x,q_y,t=0)$ in \textbf{c} with $\gamma_\downarrow=0.0, 0.4, 1.0, 2.1E_r$, respectively. }\label{FigS4}
	\end{figure}
	
	\begin{figure}[htbp]
		\centering
		\includegraphics[width=0.85\columnwidth]{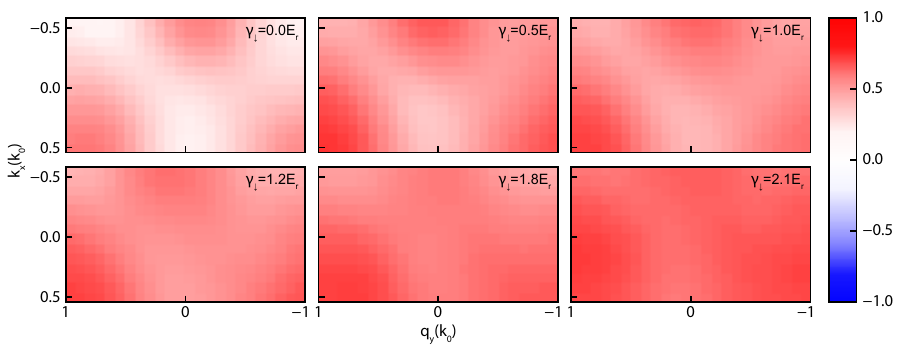}
		\caption{2D time averaged spin polarization texture obtained by the evolution from 0 ms to 0.30 ms with $\gamma_\downarrow=0.0, 0.5, 1.0, 1.2, 1.8, 2.1E_r$, respectively.}\label{FigS5}
	\end{figure}

	\subsection{Momentum-dependent time evolution}
	
	To obtain the spin information in the quasi-momentum space, we take a series of spin resolved absorption images $\{\mathcal{I}_1 (\text{no atoms killed}),\mathcal{I}_2 (\ket{\uparrow} \text{atoms killed}), \mathcal{I}_3(\ket{\uparrow,\downarrow} \text{atoms killed})\}$ after a 10ms or 15ms time-of-flight expansion. The atomic distribution of spin up and spin down can be extracted from $\mathcal{I}_\uparrow(k'_x,k_y)=\mathcal{I}_1-\mathcal{I}_2$ and $\mathcal{I}_\downarrow(k'_x,k_y)=\mathcal{I}_2-\mathcal{I}_3$. Notably, a fraction of atoms may occupy other spin states due to the imperfect isolation of the spin-$\frac{1}{2}$ subspace or spontaneous emission of the atoms on excited states pumped by the loss beam, which can be eliminated by the background image $\mathcal{I}_3$. The atomic distributions $\mathcal{I}_\sigma(k'_x,k_y)$ are then folded into the first Brillouin zone by shifting the second and third Brillouin zone with integer number of $2k_0$, $\mathcal{D}_\sigma(k'_x,q'_y)=\sum_{n=-1}^1\mathcal{I}_\sigma(k'_x,q'_y+2nk_0)$. After that, based on the spin-momentum locking, momentum distributions in quasi-momentum are defined as $\mathcal{D}_\uparrow(k_x,q_y)=\mathcal{D}_\uparrow(k'_x,q'_y)$ and $\mathcal{D}_\downarrow(k_x,q_y)=\mathcal{D}_\downarrow(k'_x+k_0\sin{\theta},\text{mod}{(q'_y+k_0-k_0\cos{\theta},2k_0)})$, where mod is the modulo operator. Finally, the spin polarization can be calculated by $P(k_x,q_y)=\frac{\mathcal{D}_\uparrow(k_x,q_y)-\mathcal{D}_\downarrow(k_x,q_y)}{\mathcal{D}_\uparrow(k_x,q_y)+\mathcal{D}_\downarrow(k_x,q_y)}$. The time evolution of spin polarization texture, normalized spin up texture and spin down texture are shown in Figure S\ref{FigS4}. In the Hermitian regime with $\gamma_\downarrow=0E_r$, the spin polarization at different quasi momentum oscillates with different frequency following Rabi process after quenching the two photon detuning and a line structure appears at specific time (0.08$\sim$0.20ms), which reveals a non-uniform band gap. When the spin-dependent atomic loss is induced, the oscillation frequency at quasi-momentum near the line structure become slower and the oscillation amplitude becomes smaller, which indicates the band gap closing behavior. The line structure, where the Raman coupling is resonant, can also be found in normalized spin up and spin down texture as shown in Figure S\ref{FigS4}b-c. In particular, when the dissipation is comparable with the Raman coupling, since the spin up atoms near the line structure are constantly transfered to the spin down states by the two photon Raman process and the spin down atoms experienced a large atomic loss, there are few atoms return spin up state, which makes the line structure in spin up texture (Figure S\ref{FigS4}b) can be observed for longer duration than Hermitian case. We can further obtain the time averaged spin texture by averaging the data with evolution time from 0 ms to 0.30 ms after quenching $m_z(\delta)$ as shown in Figure S\ref{FigS5}, where the zero time-averaged spin polarizations indicate the band inversion surface in Hermitian regime\cite{zhang2019dy, yi2019ob, song2019ob}.
	
	\begin{figure}[htbp]
		\centering
		\includegraphics[width=0.9\columnwidth]{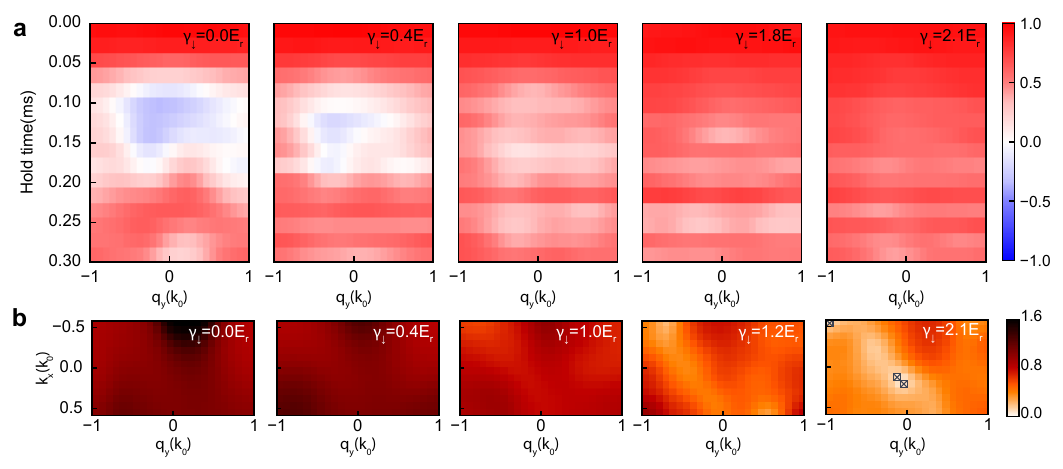}
		\caption{\textbf{a}, Time evolution of spin polarization for different loss rate at $k_x = 0$ layer. \textbf{b}, The 2D band gap reconstructed from the spin polarization texture at different dissipation strength. When the loss rate is closed to the exceptional point, the significantly reduced atom number makes it hard to obtain reasonable fitting result for some quasi-momentum, which is marked by the black cross.}\label{FigS6}
	\end{figure}

	\subsection{Extraction of band gap information}
	
	In Figure S\ref{FigS6}a, the time dependent spin texture at $k_x=0$ for different dissipation strength is shown. From the time-dependent spin polarization curve, one can roughly extract the band gap information with a empirical damped sinusoidal function $f(t)=Ae^{-Bt}\cos{Ct}+1-A$, where B corresponds to the imaginary part of energy gap and C corresponds to the real part of energy band. The extracted 2D band gap are shown in Figure S\ref{FigS6}b, which shows similar ring-like pattern as simulation from plane wave expansion in Fig.2a. 
	
	\begin{figure}[htbp]
		\centering
		\includegraphics[width=0.65\columnwidth]{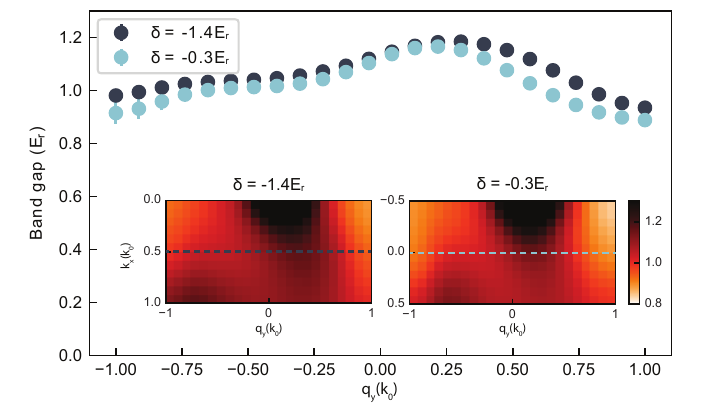}
		\caption{Comparison between reconstructed band gaps for $\delta=-0.3E_r$ (inset on the right) and $\delta=-1.4E_r$(inset on the left). For $\delta=-1.4E_r$, the data is further shifted along $k_x$ direction during the analysis to compensate the difference of two photon detuning. The error bars stand for the fitting error. }\label{FigS8}
	\end{figure}
	
	However, the phase diagram obtained from this empirical method deviates from the theoretical simulation, as shown in the inset of Fig.3a. Even prior to the exceptional point, a finite damping rate exists, and the oscillation frequency obtained from fitting cannot be completely zero or may have a large fitting error after exceptional point. There are three primary reasons why the phase diagram deviates from the simulation. Firstly, in addition to the uncertainty of atom loss caused by the power fluctuation of the loss beam (green shaded region in main Fig.3a), there is also momentum uncertainty induced by the finite resolution of the imaging system and the fluctuation of two-photon detuning in the system. This uncertainty causes the oscillation curve to dephase even without spin-dependent dissipation. In main Fig.3, the effect of this type of momentum uncertainty on the phase diagram is labeled by the brown shaded region and  explain the reason that the imaginary band gaps obtained by empirical damped sinusoidal function appear before the exceptional points. Secondly, when the dissipation is comparable to exceptional points, the atom number is significantly reduced, and the oscillation curve becomes a monotonic decay curve dominated by dissipation, making it difficult to obtain the oscillation term and further determine the position of the exceptional point. Finally, the theoretical time evolution is not a perfectly damped oscillation, which also contributes to the deviation.
	
	\begin{figure}[htbp]
		\centering
		\includegraphics[width=0.78\columnwidth]{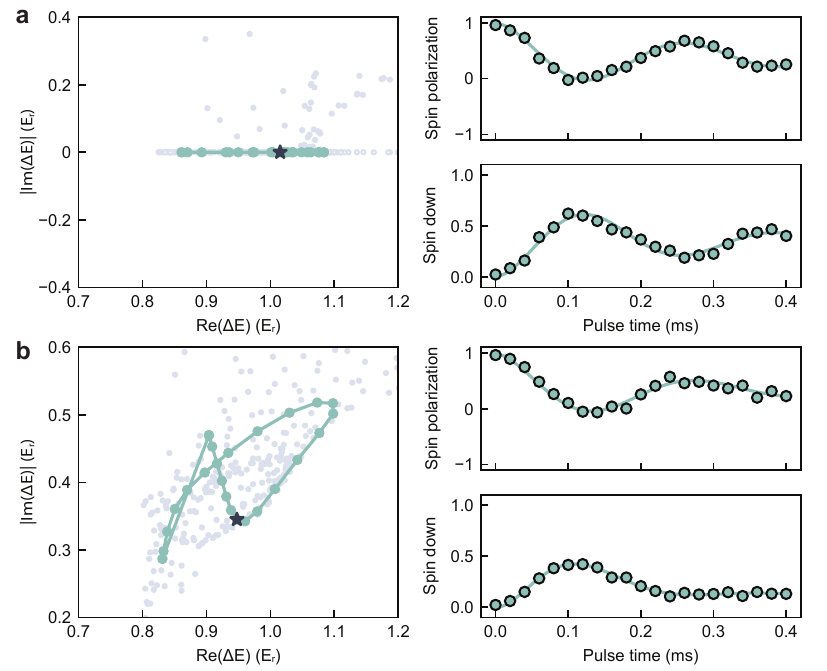}
		\caption{Reconstructing the band difference on the complex plane. \textbf{a}, Time evolution of spin polarization and normalized spin down for selected point on the complex eigen spectrum with $\gamma_\downarrow=0.0E_r$. \textbf{b}, Time evolution of spin polarization and normalized spin down for selected point on the complex eigen spectrum with $\gamma_\downarrow=0.4E_r$. In both figures, the dots are experiment data and the solid line is the fitting result.  }\label{FigS9}
	\end{figure}
	
	Therefore, we also quantify the different behaviors before and after PT-symmetry breaking transition by fitting with both damped oscillation $f(t)=Ae^{-Bt}\cos{Ct}+1-A$ and exponential decay $f(t)=Ae^{-Bt}+1-A$, and then compare the $R^2=1-\frac{\sum_{i=1}^{n}(y_i-f(x_i))^2}{\sum_{i=1}^{n}(y_i-\bar{y})^2}$, like shown in main Fig.3b.  When the loss rate is large enough, the $R^2$ of both fits overlap with each other, which indicates the needless of the oscillation term when fitting the time evolution at large loss. With this method, we can estimate where the transition occurs and then fit the time evolution curve before and after transition region with sinusoidal function and exponential function respectively to extract the energy band and damping rate (Fig 3a). For different $k_x/k_0$ (or equivalently $\delta$), the loss rate of EPs are different.
	
	\subsection{Relation between two photon detuning and $k_x$}
	
	From the Hamiltonian of our Raman lattice, one can see that although the lattice potential is one-dimensional in the $y$ direction, the running-wave term $e^{ik_2x}$ in the Raman potential couples the spin-up (spin-down) states with the x-directional kinetic energy $\frac{\hbar^2}{2M}(k_x\pm\frac{k_2}{2})^2$. This coupling gives rise to a linear SO term $\frac{\hbar^2}{2M}(k_xk_2)\sigma_z$, which leads to a 2D energy band. In fact, the linear SO term effectively shifts the Zeeman energy $m_z$, making the energy band for the $k_x=0$ layer obtained by tuning $m_z(\delta)$ identical to that for fixed $m_z(\delta)=m_0$ by scanning $k_x$ following the relation $H_{q_y,k_x=0,m_z=m_0}=H_{q_y,k_x=\frac{2Mm_0}{\hbar^2k_2},m_z=0}$.
	
	In main Fig. 3a and b, we chose different detuning values and then transferred them into different $k_x$ values due to the limited atom number at high $k_x$. In Figure S8, we compared the reconstructed band gap by shifting the two-photon detuning values $\delta$ and $k_x$ in Hermitian regime, which confirmed the relationship between the two-photon detuning and $k_x$ in our experiments.

	\subsection{Momentum uncertainty and reconstruction of band difference on complex plane}
	
	As mentioned in Section F, the momentum uncertainty makes it hard to distinguish dephasing issue from the decay induced by atom dissipation when fitting the oscillation of spin polarization with damped oscillation.  Therefore, to avoid the effect of dephasing issue, when we reconstruct the complex eigen spectrum in low dissipation regime, the imaginary gap is determined by fitting the oscillation curve of spin down atoms by following equation    
	\begin{equation}
		f(t)=A(e^{2\Im\left(E_2\right)t}+e^{2\Im\left(E_1\right)t})
		-2B{e}^{\Im\left({E_1+E}_2\right)t}\cos{\Re\left(E_2-E_1\right)t}+2B-2A
	\end{equation}  
	This equation is obtained by combining the time-dependent SchrÃ¶dinger equation of tight-binding Hamiltonian and an empirical damped sinusoidal function. Here, we use the parameters obtained from fitting the spin polarization with damped sinusoidal function as the initial fitting parameters. This fitting function enables us to distinguish the decay induced by the imaginary band from the dephasing induced by the momentum uncertainty. By combining the real band gap extracted from the oscillation curve of the spin polarization, we can further reconstruct the band difference on the complex plane. Figure S\ref{FigS9} displays the time evolution of spin polarization and normalized spin down for selected points on the complex eigen spectrum with $\gamma_\downarrow=0.0E_r$ and $\gamma_\downarrow=0.4E_r$, respectively, along with the fitting curve. When there is no spin-dependent dissipation, the normalized spin down curve with only dephasing induced by momentum uncertainty has a non-zero offset after a sufficiently long time. This behavior can be described by the above fitting function with $\Im\left(E_1\right)=\Im\left(E_2\right)$ and $A=0$, which is an empirical damped sinusoidal function. Conversely, when $\gamma_\downarrow\ne0$, the spin down atoms eventually reach zero, which implies that $\Im\left(E_1\right)\ne\Im\left(E_2\right)$ and $A=B$ in the above equation, which is the exact solution of time-dependent SchrÃ¶dinger equation of tight-binding Hamiltonian.
	
	\begin{figure}[htbp]
		\centering
		\includegraphics[width=0.85\columnwidth]{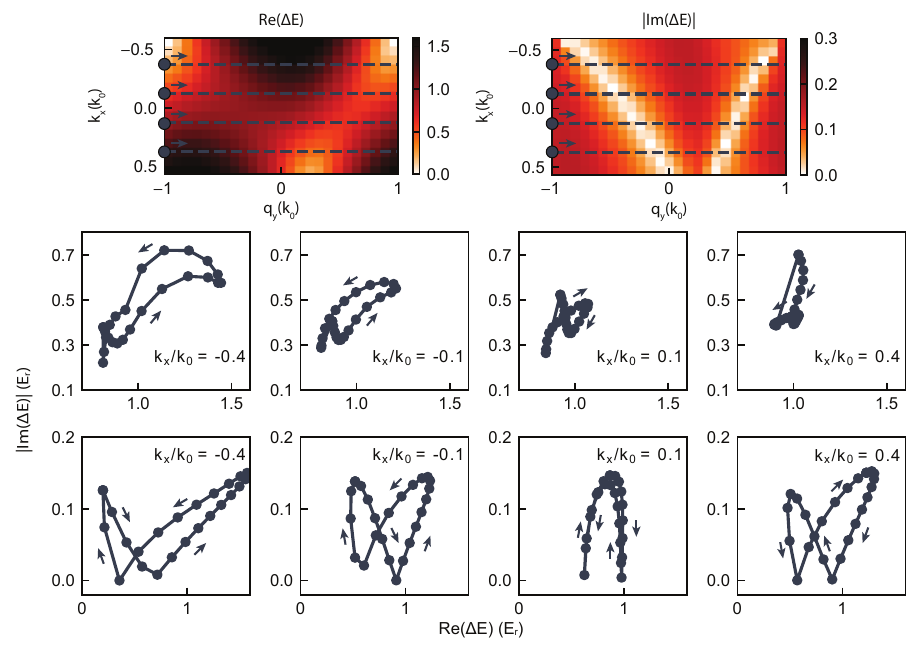}
		\caption{Band difference at different $k_x$ layers on complex plane with $\gamma_\downarrow=0.4E_r$. The top row is the experimental data, while the bottom row is the corresponding simulation result.}\label{FigS12}
	\end{figure}

	\subsection{Band difference on complex plane}
	
	Figure S\ref{FigS12} presents the band difference on the complex plane at various $k_x$ layers ($k_x/k_0=-0.4, -0.1, 0.1, 0.4$) with dissipation $\gamma_\downarrow=0.4E_r$. For all the $k_x$ layers depicted in Figure S\ref{FigS12}, the band difference exhibits a non-zero spectral winding number on the complex plane. This non-backstepping curve signifies that the system maintains a universal non-Hermitian skin effect (NHSE) along the $y$ direction, as corroborated in section II.B. It is also important to highlight that only the absolute value of the imaginary band gap can be measured in the experiment. According to the simulation, the original band difference spectrum on the complex plane manifests a ring-like pattern, and the imaginary part of the band gap alters its sign when the quasi-momentum traverses the band inversion surface. By flipping the portion with a negative imaginary band gap into the first quadrant, the ring-like pattern transforms into the butterfly shape depicted in the figures. In the case along the $q_y$ direction, as no optical lattice exists in this direction, the band difference spectrum intersects at $k_x\rightarrow\pm\infty$. Since only a portion of the band difference spectrum is measured in the experiment, it does not display a non-zero spectral winding. However, for specific $k_x$ layers ($q_y/k_0=-0.2, 0.6$), the currently measured region is sufficiently expansive to observe the non-backstepping behavior. Consequently, the system also upholds a universal NHSE along the $x$ direction, and based on these two points, the entire 2D system must exhibit a non-zero spectral area.
	
	\begin{figure}[htbp]
		\centering
		\includegraphics[width=0.85\columnwidth]{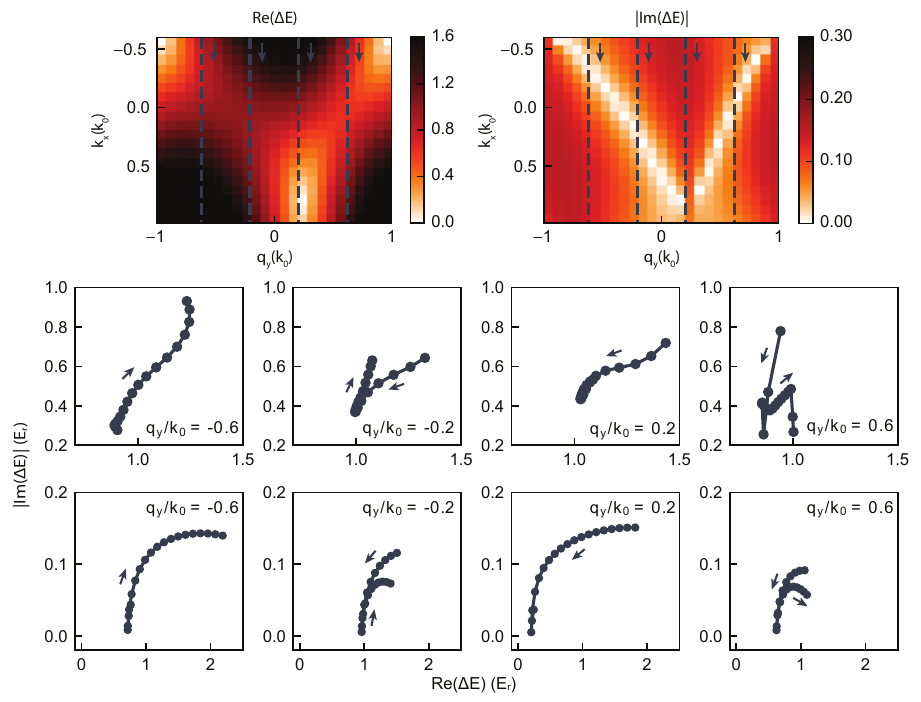}
		\caption{Band difference at different $q_y$ layers on complex plane with $\gamma_\downarrow=0.4E_r$. The top row is the experimental data, while the bottom row is the corresponding simulation result.}\label{FigS13}
	\end{figure}
	
	\red{\subsection{Real space measurement with initial motion along -x direction}}
	
	\red{By initializing the motion along $-x$ direction, in the presence of skin effect in the +x boundary ($\gamma_\downarrow=0.2E_r$), the atom cloud stopped the motion to $-x$ direction in a relatively short time and then turned to $+x$ direction. In contrast, without the skin effect ($\gamma_\downarrow=0.0E_r$), the atom cloud can move to the $-x$ direction with a larger displacement and exhibit dipole oscillation. The results are also well consistent with the numerical results (Fig.~\ref{FigS16}), further confirm the existence of skin effect with skin modes mainly accumulated to the +x direction in our present system.\\}
	
	\begin{figure}[htbp]
		\centering
		\includegraphics[width=0.65\columnwidth]{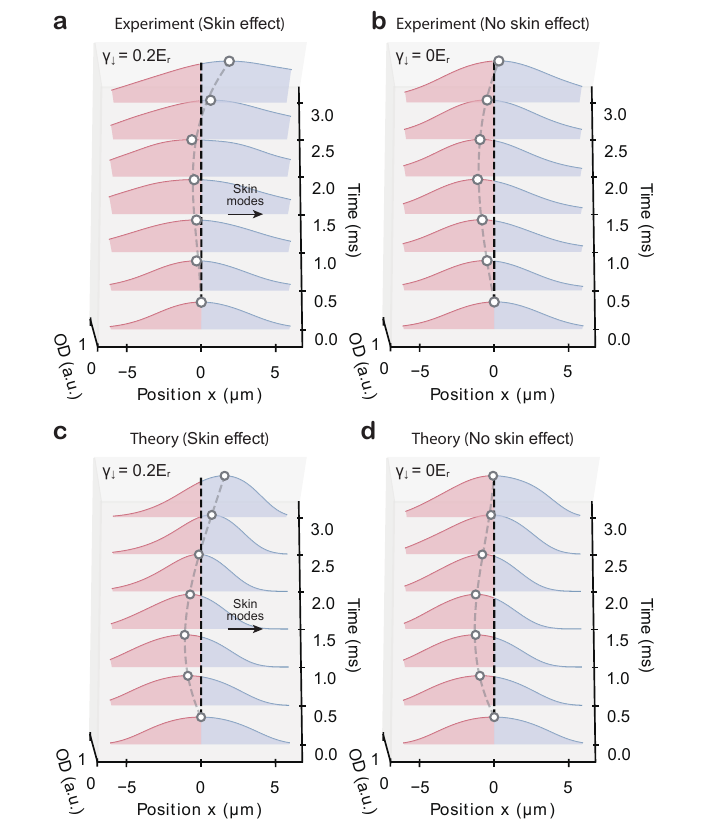}
		\caption{\red{\textbf{Real space measurement with initial motion along $-x$ direction.} \textbf{a,b} One example of experimental one-dimensional normalized spin up atom profiles along the $x$ direction with ($\gamma_{\downarrow}=0.2E_r$) or without skin effect ($\gamma_{\downarrow}=0.0E_r$). The initial motion is along $-x$ direction. The normalized atom profiles are averaged for the whole atom cloud along $y$ direction. The positions with the highest optical density (OD) are denoted by empty circles and the gray dashed line is a guideline based on polynomial function. The profiles corresponding to positive and negative positions are indicated by blue and red colors, respectively. \textbf{c,d} Simulated one-dimensional normalized spin up atom profiles along the $x$ direction with or without skin effect, specifically $\gamma_{\downarrow}=0.2E_r$, or $\gamma_{\downarrow}=0.0E_r$. The initial motion is along $-x$ direction. The normalized atom profiles are averaged for the whole atom cloud along $y$ direction. The positions with the highest optical density (OD) are denoted by empty circles and the gray dashed line is a guideline based on polynomial function. The profiles corresponding to positive and negative positions are indicated by blue and red colors, respectively. }}\label{FigS16}
	\end{figure}
	
	\section{Spectral topology and non-Hermitian skin effect}
	
	In this section, we discuss the complex spectral topology and the NHSE in detail. In the first subsection we review the theorems for skin effect in lattice systems and generalize the theory to continuous space. In the second subsection, based on the band difference obtained in the experiment we get access to non-trivial spectral topology and further predict the NHSE under open boundary condition (OBC). In the third subsection we numerically solve a 2D lattice model and discuss the features of the skin effect from a symmetry point of view. And in the last subsection we map the low-energy sectors of our 2D non-Hermitian model to Hermitian systems of free particles living in 4D curved space, and compare the non-Hermitian skin effect with the accumulation of eigenstates on the curved space.
	
	\subsection{The criteria for non-Hermitian skin effect in 1D and 2D systems}
	
	In this subsection we review the theorems for the occurrence of the skin effect in (quasi-)1D and 2D lattice system and generalize the theory to the case of continuous space. The NHSE has a topological origin manifested by the bulk-boundary correspondence between the spectral topology of the periodic boundary eigenspectrum and the skin modes in the real space. For (quasi-)1D lattices, it has been rigorously proved that the emergence of skin effect is equivalent to the existence of nonzero spectral winding numbers
	\begin{equation}
		w(E_0)=\int_{\rm{FBZ}} \frac{dk}{2\pi i}\partial_{k}\det(H(k)-E_0),
	\end{equation}
	where $H(k)$ is the Bloch Hamiltonian with quasi-momentum $k$, and $E_0$ could be any gap point~\cite{sato2020or}. An equivalent and more intuitive description is that the periodic boundary spectrum $E_i(k)$ ($i$ runs over all bands) is non-backstepping on the complex plane. Here we make a generalization that this also applies for the skin effect in the continuous space. This is because for any given eigenstate of a Hamiltonian in the continuous space we could treat it as an eigenstate of a lattice model by applying the finite difference approximation with a proper step length. The discretized model obtained in this way serves as a good approximation for the low-energy and long-wavelength spectrum of the continuous model. So if the spectrum of the continuous model is non-backstepping, that of the lattice model should also be so and the corresponding open boundary eigenstates of the lattice model should be skin modes. Therefore the continuous model should also have skin modes and exhibit NHSE. We try this criterion on two continuous two-band models which in the momentum space are written as:
	\begin{align}
		H_1(k)&=\begin{pmatrix}
			\hbar^2(k-Q)^2/2m & \Omega_R\\
			\Omega_R^* & \hbar^2(k+Q)^2/2m-i\Gamma
		\end{pmatrix},\nonumber\\
		H_2(k)&=\begin{pmatrix}
			\hbar^2(k-Q)^2/2m & \Omega_R+\Gamma/2\\
			\Omega_R^*-\Gamma/2 & \hbar^2(k+Q)^2/2m
		\end{pmatrix},\label{1Dmodels}
	\end{align}
	where $Q,\Omega_R$ and $\gamma$ are all constants. If $Q$ takes the value $k_0 \sin76^o/2$ and $\Gamma=\gamma_\downarrow/2$ is the loss rate of spin down atoms, $H_1$ is just the quasi-1D layer of our experimental model. We could see in Fig.~\ref{FigS911}(a1,a2) that $H_1$ has non-backstepping periodic boundary spectrum and exhibits skin effect under the OBC, while $H_2$ has backstepping spectrum and no skin effect.
	
	In a 2D periodic system, a nonzero periodic boundary spectral area on the complex plane results in the universal skin effect, which means that the skin effect persists unless the open boundary geometry coincides with some spatial symmetry of the bulk~\cite{zhang2022un}. And more specifically in a 2D lattice with a rectangular open boundary, the skin effect can be absent only if the spectral winding numbers of quasi-1D layers perpendicular to the boundary
	\begin{equation}
		w_E(\mathbf{k}_\|)=\int_{\rm{FBZ}} \frac{dk_\perp}{2\pi i}\partial_{k_\perp}\det(H(k_\perp,\mathbf{k}_\|)-E_0)
	\end{equation}
	are zero for all gap points $E_0$ and fixed momenta $\mathbf{k}_\|$ parallel to the boundary. Such winding numbers indicate the direction of spectral flow and on which boundary the skin modes accumulate in the real space. If the system is continuous space in one or two directions, the spectral winding numbers are no longer available. In that case we just need to check whether the spectrum of the quasi-1D layers are non-backstepping and the spectral flow will be clear. 
	
	\begin{figure}[b]
		\centering
		\includegraphics[width=0.9\columnwidth]{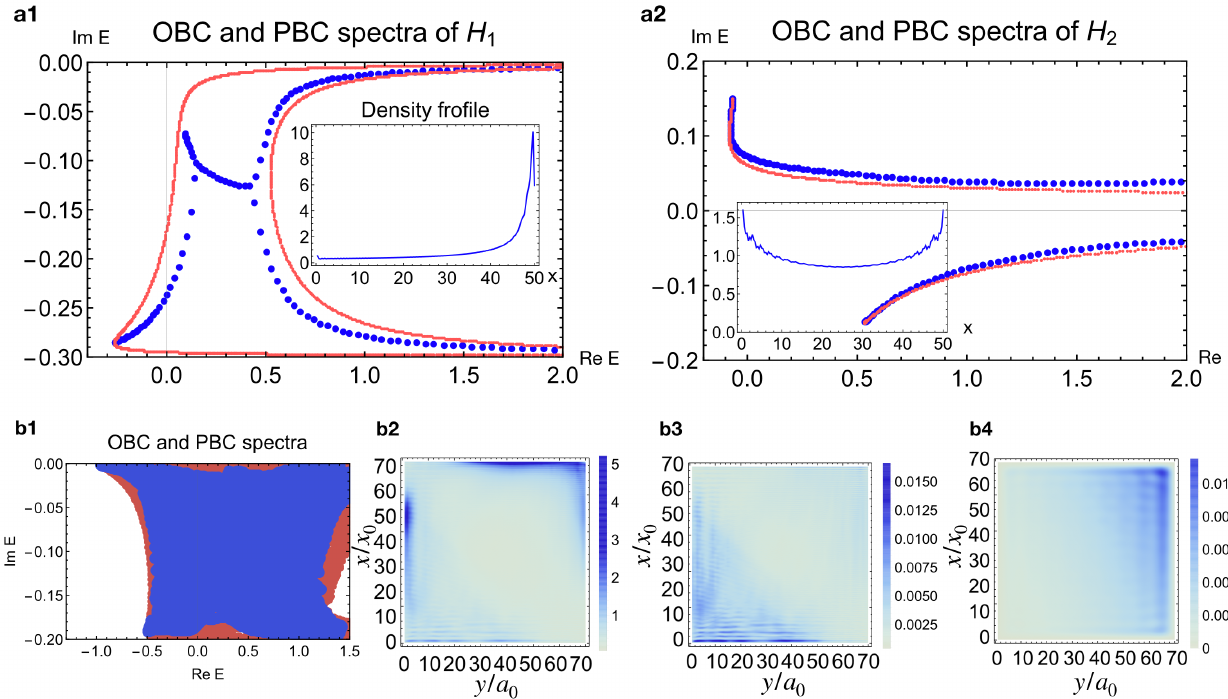}
		\caption{Complex spectra and non-Hermitian skin effect of the 1D and 2D models. \textbf{a1,a2}, open boundary spectrum (blue) and periodic boundary spectrum (red, calculated in the momentum space) of $H_{1/2}$ in Eq.~\ref{1Dmodels}. The insets are the density profile summed over all open boundary eigenstates with eigenenergy $E$ satisfying $|\Re E|<2\rm{Er}$ and spin degree of freedom  $\sum_{i,s,|\Re E_i|<2\rm{Er}} |\Psi_i(x,s)|^2$. \red{\textbf{b1}, low energy part of the periodic boundary (red) and open boundary (blue) spectra of the 2D lattice model in Eq.~\eqref{2Dlattice}. \textbf{b2}, the density profile summed over open boundary eigenstates of the model in Eq.~\ref{2Dlattice} with eigenenergy $E$ satisfying $\Re E<1.5\rm{Er}$.  \textbf{b3}, density profile of the 100 eigenstates with eigenenergies most closest to the base point $E_0=(1.0-0.2 \text{I})E_r$. \textbf{b4}, density profile of the 100 eigenstates with eigenenergies most closest to the base point $-0.2\text{I}-E_0$.}}\label{Fig.NHSE}
	\end{figure}
	
	\subsection{A proof of the nontrivial spectral topology and the prediction of skin effect}
	
	In this subsection we deduce the nontrivial spectral based on the experimental results and predict the occurrance of NHSE under OBC. The band difference $\Delta E(k_x,q_y)=E_1(k_x,q_y)-E_2(k_x,q_y)$ between the two lowest bands extracted from the quench dynamics of spin polarization shows nonzero spectral area and non-backstepping curves of quasi-1D layers in $k_x$ and $q_y$ directions. But to predict skin effect we need the actual band dispersion rather than their difference. To bridge the gap, here we show that there exists at least one topological nontrivial quasi-1D subsystem in both $x$ and $y$ directions. We make the following assertion and provide a proof: If the band difference $\Delta E(k_x,q_y)$ of the quasi-1D layers with fixed $k_x$ or $q_y$ is non-backstepping on the complex plane and does not pass the origin point ($\Delta E\neq 0$), the bands $E_{1/2}(k_x,q_y)$ should also be non-backstepping.   
	\begin{proof}
		We prove the case of quasi-1D layer along $q_y$ direction with fixed $k_{x0}$. First we define $h(k_x,q_y)=(E_1(k_x,q_y)+E_2(k_x,q_y))/2$, and then the bands are expressed as   
		\begin{eqnarray}
			E_{1/2}(k_x,q_y)=h(k_x,q_y)\pm \Delta E(k_x,q_y)/2. 
		\end{eqnarray}
		The bands $E_{1/2}(k_{x0},q_y)$ could be backstepping only in two cases: (i) $\forall q_y$, $\exists q_y\neq q_y',q_y''$ s.t. $E_1(k_{x0},q_y)=E_1(k_{x0},q_y')$, $E_2(k_{x0},q_y)=E_2(k_{x0},q_y'')$, or (ii) $\forall q_y$, $\exists$ only one $q_y'\neq q_y$ s.t. $E_1(k_{x0},q_y)=E_2(k_{x0},q_y')$. In case (i) there should be $q_y'=q_y''$ because such band degeneracy should be caused by a certain symmetry transforming the momentum in a physical sense, like $q_y\rightarrow -q_y$, $q_y\rightarrow \pi-q_y$ or so on. Then we equivalently have $h(k_{x0},q_y)\pm \Delta E(k_{x0},q_y)=h(k_{x0},q_y')\pm \Delta E(k_{x0},q_y')$ and thus $\Delta E(k_{x0},q_y)=\Delta E(k_{x0},q_y')$. This $q_y\rightarrow q_y'$ relation actually gives a local homeomorphism $q_y'(q_y)$ from the first Brillouin zone (FBZ) to itself. If the bands $E_{1/2}(k_{x0},q_y)$ are backstepping, $q_y'(q_y)$ and $q_y$ should go in the opposite directions ($q_y\rightarrow q_y'$ have a negative mapping degree, rigorously speaking). Therefore $\Delta E(k_{x0},q_y)=\Delta E(k_{x0},q_y')$ indicates that the $\Delta E(k_{x0},q_y)$ curve is backstepping. This contradiction excludes the possibility of the first case. In case (ii) we have $h(k_{x0},q_y)+ \Delta E(k_{x0},q_y)=h(k_{x0},q_y')-\Delta E(k_{x0},q_y')$. Since $q_y$ and $q_y'$ run in the opposite directions, there must be some $q_{y1}$ s.t. $q_{y1}=q_{y1}'$. Then we have $\Delta E(k_{x0},q_{y1})=0$, which contradicts the condition that the curve of $\Delta E$ does not pass the origin point. By excluding the possibilities of these two cases, we conclude that the two bands $E_{1/2}(k_x,q_y)=$ themselves are non-backstepping. As a sidenote, our proof applies to the half-integer winding number case that across the Brillouin zone the two bands exchange eigenstates rather than returning to the original eigenstates respectively.
	\end{proof} 
	Applying this on the non-backstepping $\Delta E(k_x,q_y)$ curves as shown by Fig.~\ref{FigS12} and Fig.~\ref{FigS13}, we deduce that the corresponding quasi-1D band spectra are non-backstepping. Combining this with the lack of reciprocity and spatial symmetry of our system, we predict the appearance of corner skin modes under the OBC, which is supported by the numerical simulation shown in Fig.~\ref{Fig.NHSE}b2. 
	
	With the non-backstepping spectrum of both quasi-1D layers in $k_x$ and $q_y$ directions, we could further infer that the periodic boundary spectrum of the whole 2D system must have a nonzero spectral area. This is because otherwise the spectrum on the complex plane should be a 1D curve including both non-backstepping open curves of $k_x$ layers going to real positive infinity and the non-backstepping loops of $q_y$ layers, which is not possible. The nonzero spectral area also supports the occurrence of 2D NHSE~\cite{zhang2022un}.

	\subsection{Simulation of the non-Hermitian skin effect and the nonlocal feature}
	
	To see the NHSE of our optical lattice system, here we solve a 2D lattice model with open boundary cut along $x$ and $y$ directions. \red{The harmonic trap in the experiment is replaced by the hard box trap so that the skin modes will be localized on the edges.} The model is obtained by applying s-band tight-binding approximation in the $y$ direction and finite-difference approximation in the $x$ direction. It acts on the local basis $|\mathbf{j}s\rangle$ centered at the site $\mathbf{j}=(j_x,j_y)$ with spin $s=(\uparrow,\downarrow)$, which in the $y$ direction are the s-band Wannier functions. The Hamiltonian is
	\red{\begin{align}
			H_{lattice}=&\sum_{\mathbf{j}}\Big\{(m_z+i\gamma_\downarrow/4)\bigr(|\mathbf{j}\uparrow\rangle\langle\mathbf{j}\uparrow|-|\mathbf{j}\downarrow\rangle\langle\mathbf{j}\downarrow|\bigr)        +\sum_{k=x,y}\Big[\bigr(-t_{\uparrow}^k|\mathbf{j}+\mathbf{e}_k\uparrow\rangle\langle\mathbf{j}\uparrow|+t_\downarrow^k|\mathbf{j}+\mathbf{e}_k\downarrow\rangle\langle\mathbf{j}\downarrow|\bigr)+h.c.\Big]\nonumber\\
			&+\bigr(t_{so}e^{-ik_0 \cos\theta j_y}|\mathbf{j}\downarrow\rangle\langle\mathbf{j}+\mathbf{e}_y\uparrow|-t_{so}e^{ik_0 \cos\theta j_y}|\mathbf{j}\uparrow\rangle\langle\mathbf{j}+\mathbf{e}_y\downarrow|+h.c.\bigr)\Big\}-i\gamma_\downarrow/4,\label{2Dlattice}
	\end{align}}
	where $m_z$ is the Zeeman energy, $\gamma_\downarrow$ is the loss rate of spin-down atoms, $t_{\uparrow,\downarrow}^k$ are the nearest-neighbor spin-conserved hopping coefficients of the optical lattice in the $k=(x,y)$ directions, $t_{so}$ is the nearest-neighbor spin-flipped hopping coefficient in the optical lattice direction, and $\theta=76^o, k_0=\pi/a$ with $a$ being the lattice constant in the y direction. \red{Here $t_{\uparrow}^x=t_{\downarrow}^x=i \sin\theta \frac{a}{2l_x \pi}+\frac{a^2}{l_x^2 \pi^2}$ where $l_x$ is the finite difference step-length. We could gauge out the phase factor $e^{\pm i k_0 \cos\theta j_y}$ with a transformation $|\mathbf{j}\downarrow\rangle\rightarrow e^{i k_0 \cos\theta j_y}|\mathbf{j}\downarrow\rangle$.} Here the tight-binding parameters are set to be $t_{\uparrow(\downarrow)}=0.32(0.38)E_r$ and $t_{so}=0.25E_r$ corresponding to the experimental parameters $V_{\uparrow(\downarrow)}=3.2(2.2)E_r$ and $M_R=2.0E_r$, and the Zeeman splitting is $m_z=-0.3E_r$. While 2D lattice model may fail in the high-energy and short wavelength sectors for the absence of higher bands, long-range hoppings and the discretization of the continuous direction, it could approximate the low-energy and long wavelength sector of the original system well. Here the step-length in the $x$ direction is chosen to be $x_0=a_0/7$. Fig.~\ref{Fig.converge}a shows that in this case the periodic boundary solutions with the real part of the eigenenergies satisfying $\Re E<1.5E_r$ already converge well. More quantitatively, the convergence could be demonstrated by the number of eigenstates in this regime as shown in Fig.~\ref{Fig.converge}b. Since the low-energy solutions of the lattice model are good approximation of the real eigenstates, we could study the NHSE by solving $H_{lattice}$ under the OBC.
	
	\begin{figure}[htbp]
		\centering
		\includegraphics[width=1.0\columnwidth]{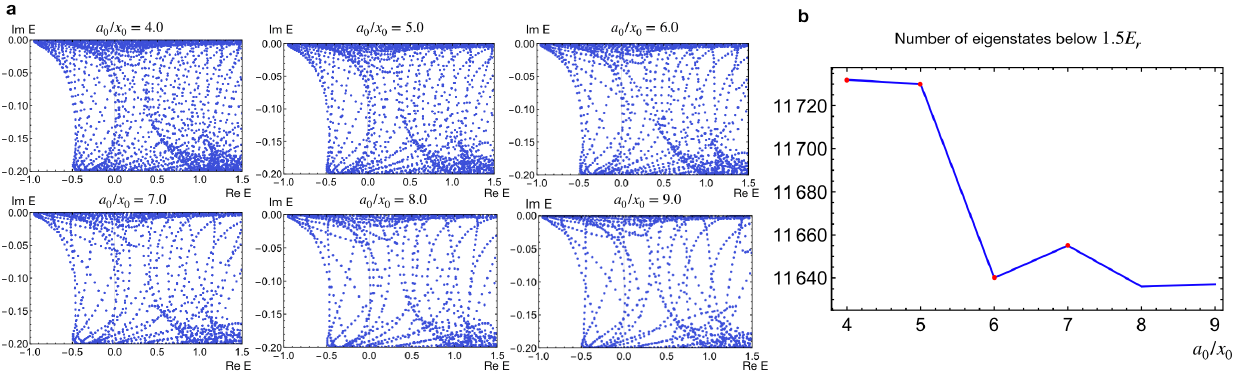}
		\caption{The convergence of the finite difference method. \textbf{a} Low energy part of the periodic boundary spectrum of the lattice model in Eq.~\eqref{2Dlattice} with different step-lengths in the $x$ direction. \textbf{b} Number of eigenstates with the real part of eigenenergy satisfying $\Re E<1.5E_r$ varying with the step-length in the $x$ direction.}\label{Fig.converge}
	\end{figure}
	
	The lattice model exhibits NHSE when $\gamma_\downarrow\neq 0$, with a volume-law number of eigenstates being localized on the corners of the open boundary as shown by Fig.~\ref{Fig.NHSE}b2. A noteworthy feature is that besides the most of the skin modes being localized at the top left corner, there are also skin modes showing up in the opposite bottom right corner. We attribute this to the (approximate) particle-hole(-like) symmetry (PHS). If $t_\uparrow^y=t_\downarrow^y$ and the overall dissipation $-i\gamma_\downarrow/4$ is omitted, the lattice model in Eq.~\eqref{2Dlattice} has a symmetry $\hat{S}=(-1)^{j_x}\sigma_x\hat{P}$, where $\sigma_x$ is a Pauli matrix acting on the spin degree of freedom and $\hat{P}$ is the transpose operation under the basis $|\mathbf{j}s\rangle$ (which is the particle-hole conjugation $\Pi_{\mathbf{j},s}(c^\dag_{\mathbf{j},s}+c_{\mathbf{j},s})$ in the second quantization representation). The measurement of $\Delta E(k_x,q_y)$ reveals the nontrivial spectral winding topology of quasi-1D layers, say $E_1(q_y)$ with a fixed $k_{x0}$. And then the particle-hole pair related to one band $E_1(q_y)$ by $\hat{S}$, $-E_1(-q_y)$, should wind in the opposite direction. Therefore the particle and hole bands have opposite spectral winding numbers, indicating that the skin modes are distributed in different (opposite) boundaries~\cite{sato2020or}. For the same reason, in the $k_x$ direction this counter-propagating spectral flows also exist in the low-energy regime. Therefore the system hosts 2D symmetric skin effect with skin modes being localized in both of the opposite corners as shown in the plot. In fact this local PHS relates pairs of skin modes like $|E\rangle,|-E\rangle$ with opposite eigenenergies $\pm E$, which if not being topological boundaries states should be localized in opposite boundaries~\cite{liu2023PH}. Fig.~\ref{Fig.NHSE}b3 show that the 100 eigenstates with eigenenergies closest to the base point $E_0=(1.0-0.2 \text{I})E_r$ are being localized on the bottom left corner, and correspondingly the 100 eigenstates with eigenenergies most closest to $-0.2\text{I}-E_0$ are being localized at the top right corner as shown in Fig.~\ref{Fig.NHSE}b4. It is a novel phenomenon arising in the non-Hermitian physics that the local symmetry establishes a nonlocal correspondence.
	
	\begin{figure}[htbp]
		\centering
		\includegraphics[width=1.0\columnwidth]{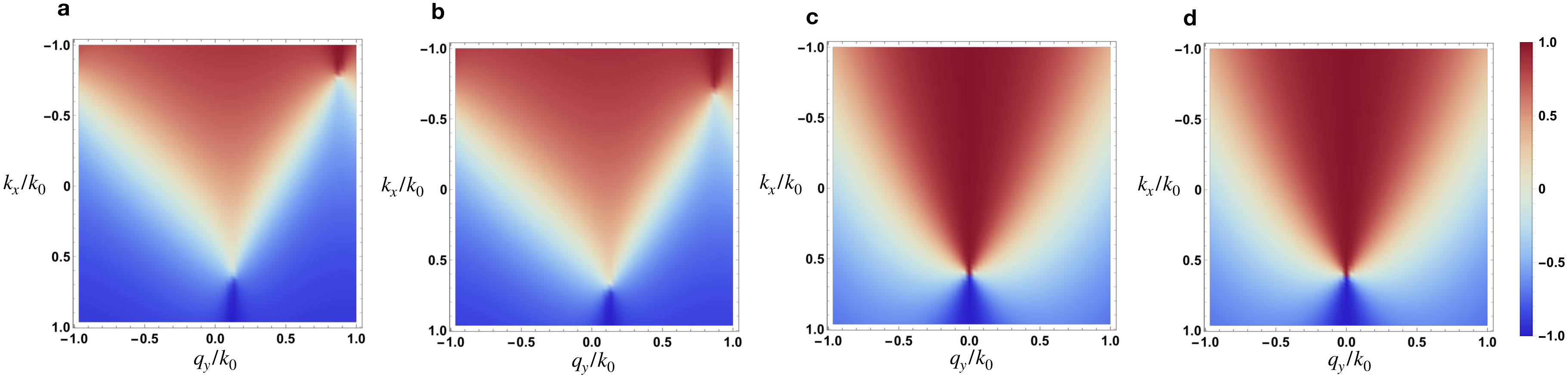}
		\caption{Numerical simulations of the spin polarization of the lowest band at equilibrium. \textbf{a} Spin polarization of the lowest band of the 2D lattice model in Eq.~\eqref{2Dlattice}, with $\gamma_\downarrow=0$ and $t_\uparrow=t_\downarrow=0.35E_r$.  \textbf{b} Spin polarization of the lowest band of the 2D lattice model in Eq.~\eqref{2Dlattice}, with $t_\uparrow=0.32Er,t_\downarrow=0.38E_r$. \textbf{c} Spin polarization of the lowest band of the model of the experimental system at equilibrium with $V_{\uparrow}=V_{\downarrow}=0.27E_r$, calculated by the plane wave method. \textbf{d} Spin polarization of the lowest band of the model of the experimental system at equilibrium with $V_{\uparrow(\downarrow)}=0.32E_r(0.22E_r)$, calculated by the plane wave method. }\label{Fig.PHS}
	\end{figure}
	
	Our real experimental system deviates from this $\hat{S}$ symmetry, mainly resulting from the high-energy sectors (higher bands beyond the energy scale of the s-band in the optical lattice) and that $t_\uparrow\neq t_\downarrow$. However, we know that the s-band width being about $w=2t_\uparrow+2t_\downarrow$ is much larger than the strength of the perturbation breaking the PHS at about $\delta w=2|t_\uparrow-t_\downarrow|$. Since any pair of eigenstates related by the PHS have opposite eigenenergies, such pair of skins modes in the low-energy sector whose eigenenergies are larger than $\delta w$ will not be largely influenced by the perturbation and their localization properties should be preserved. Therefore we could still predict this nonlocal feature of the skin effect for the real system.
	More concretely, the low-energy properties of the real system are not qualitatively changed compared with the ideal model in Eq.~\eqref{2Dlattice}. We visualise the symmetry breaking by simulating the spin polarization of the lowest band at equilibrium as shown in Fig.~\ref{Fig.PHS}: In the ideal symmetric system described by Eq.~\eqref{2Dlattice} with $t_\uparrow=t_\downarrow$, only the physics within the energy scale of s-band width of the optical lattice is kept. In this case the $\hat{S}$ symmetry is perfectly respected, reflected in the features that the Dirac points are located at $q_y/k_0=0,\pm 1$ and the reflection symmetry about the $q_y=0$ axis of the spin polarization distribution as shown in Fig.~\ref{Fig.PHS}a. If $t_\uparrow\neq t_\downarrow$, the Dirac points are slightly perturbed from the symmetric lines $q_y=0,\pm \pi$ and the reflection symmetric feature is slightly broken as shown in Fig.~\ref{Fig.PHS}b. If the higher-bands are included and the optical lattice is spin independent, the Dirac points are moved away from the symmetric lines and the spin polarization image is not perfectly symmetric about $q_y=0$ as shown by Fig.~\ref{Fig.PHS}c. In the real experimental system with spin-dependent optical lattices, the band structure is modified as shown by Fig.~\ref{Fig.PHS}d. We could see that the symmetric features protected by the PHS are broken to a small extent in the real system. Also, the simulation shown in Fig.~\ref{Fig.NHSE}b1 already shows the case of $t_\uparrow\neq t_\downarrow$ in which the nearly-symmetric feature is observed. 
	The robustness of the nonlocal skin effect with PHS significantly distinguishes PHS from time-reversal symmetries~\cite{sato2020or,cr2020,hofmann2020re} and unitary spatial symmetries~\cite{high2020,mi2020}, in the case of which the degenerate pairs related by the symmetry could be scattered to each other by an arbitrarily small perturbation breaking the symmetry, resulting in the absence of skin effect. The nonlocal nature of the fundamental symmetries like the PHS manifested in non-Hermitian physics, and the symmetry protected feature of the NHSE are the new directions of our future research. As an aside, the eigenstates with energy much higher than the s-band energy scale should recover the Hermiticity and become more extended in the real space, simply because the kinetic energy part dominates the short wavelength physics.
	
	\subsection{Mapping our 2D non-Hermitian system to Hermitian models in curved spaces}
	
	\red{In this part we study our 2D non-Hermitian experimental model under spin-orbital-coupled basis, manifesting the asymmetric hoppings and imaginary vector potential which generally induce the 2D non-Hermitian skin effect. Further, we map the low energy sectors of our model to Hermitian systems of free particles in 4D curved spaces.}

	\red{The simplified s-band Bloch Hamiltonian of our experimental model is
		\begin{align}    H(\mathbf{k})&=h_0(\mathbf{k})\sigma_0+h_z(\mathbf{k})\sigma_z+h_y(\mathbf{k})\sigma_y+i\gamma_\downarrow/4\sigma_z,\label{2DBloch}
		\end{align}
		where $\mathbf{k}=(k_x,q_y)$, $h_0(\mathbf{k})=\hbar^2(k_x^2+k_2^2/4)/2m-2t_y\cos(q_y a_0)\sin(k_1 a_0/2)-i\gamma_\downarrow/4, h_z(\mathbf{k})=-\hbar^2(k_xk_2)/2m-2t_y\sin(q_y a_0)\cos(k_1 a_0/2)+m_z, h_y(\mathbf{k})=-2t_{so}\cos(q_y a_0)$, and $\sigma_{0,y,z}$ are the Pauli matrices. Here an overall $(k_0+k_1)/2$ momentum shift has been applied in the $y$ direction. We rewrite it under the new spin-orbital-coupled basis $|\pm,\mathbf{k}\rangle$ which are the Bloch states of the Hermitian part of our model, i.e. $[H(\mathbf{k})-i\gamma_\downarrow/4\sigma_z]|\pm,\mathbf{k}\rangle=(h_0(\mathbf{k})\pm \epsilon(\mathbf{k}))|\pm,\mathbf{k}\rangle$ with $\epsilon(\mathbf{k})=\sqrt{h_z^2+h_y^2}$:
		\begin{align}
			\tilde{H}(\mathbf{k})&=h_0(\mathbf{k})\sigma_0+\epsilon(\mathbf{k})\sigma_z+i\frac{\gamma_\downarrow}{4}U^\dag(\mathbf{k})\sigma_z U(\mathbf{k})\nonumber\\
			&=h_0(\mathbf{k})+\epsilon(\mathbf{k})\sigma_z+i\frac{\gamma_\downarrow}{4}(\cos(\theta(\mathbf{k}))\sigma_z+\sin(\theta(\mathbf{k}))\sigma_x)\nonumber\\
			&=h_0(\mathbf{k})+\epsilon(\mathbf{k})\sigma_z+i\frac{\gamma_\downarrow}{4}\frac{h_z(\mathbf{k})\sigma_z+h_y(\mathbf{k})\sigma_y}{\sqrt{h_y(\mathbf{k})^2+h_z(\mathbf{k})^2}},\nonumber\\
			\text{where }U(\mathbf{k})&=\begin{pmatrix}
				\cos\frac{\theta(\mathbf{k})}{2} & \sin\frac{\theta(\mathbf{k})}{2}\\
				i\sin\frac{\theta(\mathbf{k})}{2} & -i\cos\frac{\theta(\mathbf{k})}{2}
			\end{pmatrix}, \theta(\mathbf{k})=\arctan\frac{h_y(\mathbf{k})}{h_z(\mathbf{k})}.\label{eq.rotated}
		\end{align}
		This formalism explains the occurrence of the non-Hermitian skin effect under the open boundary condition: In both $x$ and $y$ directions, the Hermitian part $h_0(\mathbf{k})+\epsilon(\mathbf{k})\sigma_z$ is parity-even and the non-Hermitian term $-i(\gamma_\downarrow h_z(\mathbf{k})\sigma_z)/(4\epsilon(\mathbf{k}))$ is parity-odd. The $-i(\gamma_\downarrow t_y\cos(k_1 a_0/2)\sin(q_y a_0)\sigma_z)/(2\epsilon(\mathbf{k}))$ term contributes imbalanced hopping amplitudes in the real space and hence induces non-Hermitian skin effect in the $y$ direction.
		Similarly in the continuous space direction the $-i(\gamma_\downarrow \hbar^2 k_2 k_x\sigma_z)/(8\epsilon(\mathbf{k})m)$ term resembles an imaginary vector potential, inducing the non-Hermitian skin effect just as how the imbalanced hoppings work in the lattice cases. Moreover, since these two terms have different signs for the two sectors $|\pm,\mathbf{k}\rangle$, the non-Hermitian skin modes should appear in two opposite corners, in agreement with the numerical results shown by Fig.~S\ref{Fig.NHSE} (b2-b4).}
	
	\red{This could be further clarified if we adopt the following approximations: (i) In the low-energy (long-wavelength) limit, i.e. $|k_x|,|q_y|\rightarrow 0$, we have $\epsilon(\mathbf{k})\approx \sqrt{m_z^2+4t_{so}^2}:=E_0$. (ii) In this regime the gap between the $|\pm,\mathbf{k}\rangle$ two sectors of $\tilde{H}(\mathbf{k})$ in Eq.~\eqref{eq.rotated}, $2|m_z|$, is much larger than the off-diagonal terms. Since the diagonal terms already manifest non-Hermitian skin effect, the $\sigma_y$ term could be neglected. The simplified model is:
		\begin{align}
			\tilde{H}(\mathbf{k})=\frac{\hbar^2 k_x^2}{2m}+t_y\sin(k_1 a_0/2)a_0^2q_y^2+\{E_0+i\frac{\gamma_\downarrow[m_z-(\hbar^2k_xk_2)/(2m)-2t_ya_0\cos(k_1a_0/2)q_y]}{4E_0}\}\sigma_z+C,
		\end{align}
		where $C$ is an overall constant and will be omitted in the following. For the $|\pm,\mathbf{k}\rangle$ bands, the diagonal terms are 
		\begin{equation}
			\tilde{H}_\pm(\mathbf{k})=\frac{\hbar^2}{2m}(k_x\mp i A_x)^2+\frac{\hbar^2}{2m_y}(q_y\mp iA_y)^2\pm (E_0+i\frac{\gamma_\downarrow m_z}{4E_0}),
		\end{equation}
		where $(A_x,A_y)=(\frac{\gamma_\downarrow k_2}{8E_0}, \frac{\gamma_\downarrow \cot(k_1a_0/2)}{4a_0 E_0})$ is the imaginary vector potential and $m_y=2 m t_y\sin(k_1a_0/2)a_0^2/\hbar^2$ is the effective mass at $q_y=0$. The imaginary vector potential of $\tilde{H}_+(\mathbf{k})$ could be removed by introducing a coordinate transformation $(\zeta,\eta)=(e^{-2A_x x},e^{-2A_y y})$:
		\begin{align}
			\tilde{H}_\pm(x,y)&=\hbar^2(-i\partial_x\mp iA_x)^2/(2m)+\hbar^2(-i\partial_y\mp iA_y)^2/(2m_y)\pm(E_0+\frac{\gamma_\downarrow m_z}{4E_0}),\nonumber\\
			\tilde{H}_\pm(\zeta,\eta)&=\hbar^2(2i A_x\zeta\partial_{\zeta}\mp iA_x)^2/(2m)+\hbar^2(2i A_y\eta\partial_{\eta}\mp iA_y)^2/(2m_y)\pm(E_0+\frac{\gamma_\downarrow m_z}{4E_0})\nonumber\\
			&=-\hbar^2A_x^2(2\zeta\partial_{\zeta}\mp 1)^2/(2m)-\hbar^2A_y^2(2\eta\partial_{\eta}\mp 1)^2/(2m_y)\pm(E_0+\frac{\gamma_\downarrow m_z}{4E_0}).
		\end{align}
		Then we have 
		\begin{align}
			\tilde{H}_+(\zeta,\eta)&=-\hbar^2A_x^2(4\zeta^2\partial_{\zeta}^2+1)/(2m)-\hbar^2A_y y^2(4\eta^2\partial_{\eta}^2+1)/(2m_y)+(E_0+\frac{\gamma_\downarrow m_z}{4E_0}),\nonumber\\
			\tilde{H}_-(\zeta,\eta)&=-\hbar^2A_x^2(4\zeta^2\partial_{\zeta}^2+8\zeta\partial_{\zeta}+1)/(2m)-\hbar^2A_y^2(4\eta^2\partial_{\eta}^2+8\eta\partial_{\eta}+1)/(2m_y)-(E_0+\frac{\gamma_\downarrow m_z}{4E_0}).\label{eq.NH}
		\end{align}
		This reminds us of the free particles living on a curved space, whose Hamiltonian is given by
		\begin{equation}
			H_{\text{cv}}=-\frac{\hbar^2}{2m}\frac{1}{\sqrt{g}}\partial_i\mathbf{g}^{ij}\sqrt{g}\partial_j,
		\end{equation}
		where $\mathbf{g}$ is the metric, $g=\det(\mathbf{g})$ and $j$ runs over the space indices. However, the Hamiltonians of the two bands, $\tilde{H}_\pm(\zeta,\eta)$, could not be simply equalized to Hamiltonians of particles living in a 2D curved space. But we can introduce two auxiliary dimensions $(\zeta',\eta')$ and map $\tilde{H}_\pm(\zeta,\eta)$ to Hermitian systems in a 4D curved space. We consider two different 4D spaces $M^\pm(\zeta,\eta,\zeta',\eta')$ with metrics
		\begin{equation}
			\mathbf{g}_\pm(\zeta,\eta,\zeta',\eta')=(\zeta^{-2}d\zeta^2+\zeta^{\mp2}d\zeta'^2)/(2A_x)^2+(\eta^{-2}d\eta^2+\eta^{\mp2}d\eta'^2)m_y/(4mA_y^2),
		\end{equation} respectively.
		And correspondingly the Hamiltonians of free particles on these two different 4D spaces are given by
		\begin{align}
			H_\text{cv}^+(\zeta,\eta,\zeta',\eta')&=-\frac{\hbar^2}{2m}\frac{1}{\sqrt{g_+}}\partial_i\mathbf{g}^{ij}_+\sqrt{g_+}\partial_j\nonumber\\
			&=-\frac{\hbar^2}{2m}4A_x^2 \zeta^2(\partial_\zeta^2+\partial_{\zeta'}^2)-\frac{\hbar^2}{2m_y}4A_y^2 \eta^2(\partial_\eta^2+\partial_{\eta'}^2),\nonumber\\
			H_\text{cv}^-(\zeta,\eta,\zeta',\eta')&=-\frac{\hbar^2}{2m}\frac{1}{\sqrt{g_-}}\partial_i\mathbf{g}^{ij}_-\sqrt{g_-}\partial_j\nonumber\\
			&=-\frac{\hbar^2}{2m}4A_x^2 (\zeta^2\partial_\zeta^2+\zeta^2\partial_{\zeta'}^2+2\zeta\partial_\zeta)-\frac{\hbar^2}{2m_y}4A_y^2 (\eta^2\partial_\eta^2+\eta^2\partial_{\eta'}^2+2\eta\partial_\eta).\label{eq.curved}
		\end{align}
		Comparing Eq.~\eqref{eq.NH} with Eq.~\eqref{eq.curved}, we notice that apart from the insignificant constant energy shifts, $\tilde{H}_\pm(\zeta,\eta)$ are exactly the Hermitian $H^\pm_{\text{cv}}$ with the quasi-momentum in the auxiliary directions $k_{\zeta'},k_{\eta'}=0$, i.e.
		\begin{equation}
			\tilde{H}_\pm(\zeta,\eta)=H^\pm_{\text{cv}}(\zeta,\eta,k_{\zeta'}=0,k_{\eta'}=0).
		\end{equation}
		These 4D spaces $M^\pm(\zeta,\eta,\zeta',\eta')$ are the Cartesian products of two curved 2D spaces parametrized by $M_\zeta^\pm(\zeta,\zeta')$ and $M_\eta^\pm(\eta,\eta')$ respectively. All of $M^\pm_{\zeta/\eta}$ have uniform negative Gaussian curvature. Taking $M^\pm_\zeta$ as an example, the Gaussian curvature is given by
		\begin{equation}
			R=-\sqrt{g_{\zeta\zeta}g_{\tilde{\zeta}\tilde{\zeta}}}^{-1}((\frac{(\sqrt{g_{\zeta\zeta}})_{\tilde{\zeta}}}{\sqrt{g_{\tilde{\zeta}\tilde{\zeta}}}})_{\tilde{\zeta}}+(\frac{(\sqrt{g_{\tilde{\zeta}\tilde{\zeta}}})_{\zeta}}{\sqrt{g_{\zeta\zeta}}})_{\zeta}).
		\end{equation}
		Simple arithmetic gives $R=-4A_x^2$ for $M^\pm_\zeta$ and $R=-(4mA_y^2)/m_y$ for $M^\pm_\eta$, which means $M^\pm_{\zeta,\eta}$ are all pseudospheres. $\mathbf{g}^+$ is exactly the standard Poincar$\text{\'e}$ upper-half-plane representation of pseudospheres, from which $\mathbf{g}^-$ differs by a transformation $(\zeta,\eta)\rightarrow(\zeta^{-1},\eta^{-1})$.}
	
	\red{The accumulation of the eigenfunctions of $H^\pm_{\text{cv}}$ in the boundary of the curved space coincides with the non-Hermitian skin effect of our 2D non-Hermitian model. The skin modes of the two sectors $\tilde{H}_\pm(\mathbf{k})$ are localized on the $(x,y)$ and $(-x,-y)$ corners respectively. While the eigenfunctions of $H^\pm_{\text{cv}}(\zeta,\eta,k_{\zeta'}=0,k_{\eta'}=0)$ are the linear superpositions of wavefunctions in the form of $\Psi(\zeta,\eta)=\zeta^{\pm 1/2+i k_\zeta}\eta^{\pm 1/2+i k_\eta}$, where $k_\zeta$ and $k_\eta$ are both real numbers. Therefore the eigenstates are localized on the $(-\zeta,-\eta)$ and $(\zeta,\eta)$ corners for $H^\pm_{\text{cv}}(\zeta,\eta,k_{\zeta'}=0,k_{\eta'}=0)$ respectively, in agreement with the distribution of non-Hermitian skin modes.}
	
	\red{It is worthwhile to note that, the 4D spaces here have two "extra dimensions" $\zeta'$ and $\eta'$ along each of which the translation symmetry is respected and the momentum is a conserved quantity. This geometry shares some similarities with the Kaluza-Klein theory~\cite{Kaluza2018On,Klein1926fi,Klein1926at}, a formalism unifying gravitation and electromagnetism. In this theory, an extra fifth dimension in the shape of a tiny circle is introduced besides the regular 3+1 space-time dimensions, and the momentum along the extra dimension is identified with the electric charge quantity. In our case, $k_{\zeta'(\eta')}$ controls the strength of the "geometrical" potential in the form of $\zeta^2$($\eta^2$) in Eq.~\eqref{eq.curved}. This inspires us to further exploit the potential of non-Hermitian lattices for the simulation of high-dimensional special curved spaces in future work.}

\end{widetext}


\end{document}